\documentclass[aps,prx,twocolumn,superscriptaddress]{revtex4-2}
\usepackage[utf8]{inputenc}
\usepackage[T1]{fontenc}
\usepackage{lmodern}
\usepackage{microtype}
\usepackage{amsmath,amssymb,amsfonts,mathtools}
\usepackage{physics}
\usepackage{bm}
\usepackage{graphicx}
\usepackage{subcaption}
\usepackage{hyperref}
\usepackage{booktabs}
\usepackage{xcolor}

\usepackage[charter,cal=cmcal,sfscaled=false]{mathdesign}
\definecolor{iffsred}{cmyk}{0.12,0.94,0.87,0.34}
\definecolor{uestcblue}{cmyk}{0.99,0.78,0.16,0.03}

\usepackage{hyperref}

\hypersetup{
  pdftitle={Local-Observable-Guided Generative Quantum Circuits for Degenerate Ground Spaces},
  pdfauthor={Chen, Zhang, Zhu et al.},
  pdfstartview=Fit,
  pdfpagelayout=SinglePage,
  colorlinks,
  linkcolor=uestcblue,
  citecolor=uestcblue,
  urlcolor=iffsred}

\begin{document}
\title{Local-Observable-Guided Generative Quantum Circuits for Degenerate Ground Spaces}

\author{Yiying Chen}
\author{Lingxia Zhang}
\affiliation{Institute of Fundamental and Frontier Sciences, University of Electronic Science and Technology of China, 611731, Chengdu, China}
\affiliation{Key Laboratory of Quantum Physics and Photonic Quantum Information (University of Electronic Science and Technology of China), Ministry of Education, 611731, Chengdu, China}
\author{Yanzheng Zhu}
\affiliation{Institute of Quantum Computing and Software, School of Computer Science and Engineering, Sun Yat-sen University, Guangzhou 510006, China}
\affiliation{Institute of Fundamental and Frontier Sciences, University of Electronic Science and Technology of China, 611731, Chengdu, China}
\author{Kaiyan Yang}
\author{Xiao Zeng}
\author{Zizhu Wang}\email{zizhu@uestc.edu.cn}
\affiliation{Institute of Fundamental and Frontier Sciences, University of Electronic Science and Technology of China, 611731, Chengdu, China}
\affiliation{Key Laboratory of Quantum Physics and Photonic Quantum Information (University of Electronic Science and Technology of China), Ministry of Education, 611731, Chengdu, China}

\begin{abstract}
Searching for degenerate ground spaces in quantum many-body systems is central to understanding spontaneous symmetry breaking and topological order. Although existing numerical methods can approximate individual ground states with high accuracy, recovering the full degenerate space remains a substantial challenge. Here we tackle this problem using a hybrid generative quantum circuit that combines a classical generative model with an expressive parameterized quantum circuit (PQC). The classical model learns a distribution over PQC parameters, enabling the sampling of an ensemble of ground states, while the PQC ensures compatibility with quantum hardware. To promote both low energy and state diversity, we define an energy-diversity objective composed of an energy-minimization term and cosine-similarity penalties derived from local observable correlators. These local descriptors provide a scalable, measurement-efficient means of distinguishing distinct ground states. We benchmark the framework on the Majumdar-Ghosh model, the Affleck-Kennedy-Lieb-Tasaki model, and the spin-1 XXZ chain, which realize distinct mechanisms of degeneracy. In all cases, the method produces a diverse ensemble whose linear span accurately reproduces the target ground space, in some instances, it identifies an approximately orthogonal basis within the learned ensemble. We further show that the framework remains robust under shot-based estimation and can still recover the degenerate ground space with a reduced measurement budget.

\end{abstract}
\maketitle

\section{Introduction}
Understanding the ground-state structure of quantum many-body Hamiltonians is a fundamental challenge in condensed matter physics, quantum information, and quantum simulation. In particular, many physical systems exhibit a degenerate ground space, arising from phenomena such as frustration, symmetry breaking, or topological order~\cite{Baumann2011_PRL,Chen2014_NatCommun,LevinWen2006_PRL,Venderbos2012_PRL,Su2013QuantumFidelity,Hamma2016MutualInfoSSB,NussinovOrtiz2009SufficientSymmetryTQO,villain1977theory,affleck1987valence}. Characterizing the complete set of degenerate ground states is crucial for diagnosing phase structure, identifying spontaneously broken symmetries, and uncovering signatures of topological or symmetry-protected order~\cite{RevModPhys.89.025003,doi:10.1126/science.aal3099}.

While much progress has been made in approximating individual ground states, representing full bases of degenerate ground states remains challenging. Conventional methods, such as exact diagonalization (ED), provide direct access to the full eigenspectrum and corresponding eigenstates, but they are limited by exponential scaling with system size~\cite{Lauchli2009_EDSmartTool,JungNoh2020_EDGuide}. Tensor-network approaches, such as Matrix Product State (MPS) and Projected Entangled Pair State (PEPS)~\cite{RevModPhys.93.045003,SCHOLLWOCK201196}, extend the accessible system sizes, yet require careful initialization or symmetry resolution to recover multiple degenerate states~\cite{He2014TopDegDMRG,Vaezi2017EntanglementDistanceDMRG,Pfeifer2015SPTLocalMinima}. More recently, the variational quantum eigensolver (VQE) and its extensions have provided a hardware-compatible route to ground-state preparation through parameterized quantum circuits (PQCs)~\cite{peruzzo2014variational,nakanishi2019subspace,mcclean2016phase,bergholm2018pennylane,qiskit2024,xu2024mindspore}. However, standard VQE formulations are designed primarily to find a single ground state rather than to explore the full degenerate space. In principle, repeated VQE runs could be used to sample multiple ground states, but this strategy is computationally costly and becomes increasingly vulnerable to barren plateaus (BP) as the circuit size grows~\cite{Hanin2018ExplodingVanishing}. Methods such as VQE-SA, which employ a small-angle initialization strategy, can alleviate gradient suppression through carefully chosen initialization~\cite{Grant2019Initialization}, but may also restrict the diversity of accessible states~\cite{zhangVariationalOptimizationQuantum2025}, so that optimization often collapses onto only a small subset of the degenerate ground space.

A promising alternative is the Variational Generative Optimization Network (VGON)~\cite{zhangVariationalOptimizationQuantum2025}, which was introduced to mitigate vanishing-gradient effects in the BP and to address degenerate-ground-state problems through a generative framework. VGON contains a pair of deep feed-forward neural networks connected by a stochastic latent layer, and a problem-specific objective function. By training on batches of random inputs, it learns a probability distribution over parameters rather than a single optimized configuration. This generative feature makes it naturally suited to many-body systems with ground-state degeneracy, where one aims to generate a family of states spanning the target degenerate space. Nevertheless, the central difficulty remains unresolved: the optimization must balance two competing objectives: drive generated states toward the ground space and maintain sufficient diversity to avoid representation collapse. Without a robust architecture and a carefully designed objective, the learned distribution may still concentrate on only a few representative states rather than covering the full space.

To overcome this difficulty, the diversity term in the objective should be both physically meaningful and computationally efficient. In particular, it should distinguish different degenerate ground states using observables that remain scalable with system size. A natural route is suggested by recent advances in reconstructing ground states of local Hamiltonians from local data~\cite{swingle2014reconstructing,huangPredictingManyProperties2020a,cramerEfficientQuantumState2010a}. It has been shown that a wide range of quantum states, such as MPS, can be faithfully reconstructed from a limited set of local measurements and classical postprocessing, with computational cost scaling polynomially with system size~\cite{cramerEfficientQuantumState2010a}. These findings indicate that suitably chosen local observables can serve as low-cost, informationally rich features that capture the essential distinctions between degenerate ground states, avoiding exponentially costly tomography.

Building on these insights, we construct a hybrid classical-quantum generative framework by combining VGON with an expressive PQC, which we refer to as a generative quantum circuit, and introduce an energy-diversity objective that simultaneously minimizes energy and promotes linear independence among generated states. In this framework, a deep classical generator produces a distribution over circuit parameters, while the PQC prepares the corresponding many-body states directly on quantum hardware. The objective is defined on a batch of generated states and combines the energy expectation with a diversity penalty based on the cosine similarity of local feature vectors constructed from one-site expectation values and two-site correlators. In this way, the generative quantum circuit is guided by efficient local data to explore a diverse family of states whose span covers the full degenerate ground space.

In this work, we study three quantum spin models with distinct mechanisms of
ground-state degeneracy: the spin-$\frac{1}{2}$ Majumdar-Ghosh (MG) model, the spin-1 Affleck-Kennedy-Lieb-Tasaki (AKLT) model, and the spin-1 XXZ chain. To cover these systems, we employ expressive circuit ans\"atze adapted to the underlying spin representations, including a symmetry-based encoding for the spin-1 setting~\cite{yangCostLocallyApproximating2025,zhangVariationalOptimizationQuantum2025}. Using the generative quantum circuits, we show that local observables can provide sufficient guidance for learning the structure of a degenerate ground space beyond isolated solutions, including cases with nontrivial structure such as symmetry-encoded models. In particular, the trained models capture the ground-space degeneracy and produce diverse ground states whose span agrees with the target ground space. In some cases, they also yield approximately orthogonal basis states. We further investigate the MG model under finite-shot estimation. Unlike existing finite-shot optimization methods, which mainly address the convergence of single-state variational objectives under limited measurements~\cite{kubler2020adaptive,zhu2024optimizing,Gu2021AdaptiveSA}, our results indicate that the generative framework remains effective in recovering the full ground space even with a reduced measurement budget.

Our main contributions are:
\begin{itemize}
    \item We introduce a local-observable-guided energy--diversity objective for training generative quantum circuits to sample from degenerate ground spaces, using cosine-similarity penalties built from one- and two-body correlators.
    \item We evaluate ground-space recovery using span-level diagnostics (tolerance rank, principal angles, and chordal distance) that quantify whether the generated ensemble covers the full degenerate manifold.
    \item We demonstrate robustness in a finite-shot setting by combining a shot-adaptive schedule with a stochastic parameter-shift update that reduces gradient-evaluation cost.
\end{itemize}
Throughout, the training objective depends only on the Hamiltonian energy and the chosen local observables. Access to the exact ground-space basis is used only for benchmark certification (overlap score $p_G$ and principal angles) on system sizes accessible to exact diagonalization.

\section{Problem setting}
\label{sec:Problem}
Given a many-body Hamiltonian $H$ defined on a finite spin chain whose ground space is degenerate, let $E_0$ denote the ground state energy, the corresponding ground space is $\mathcal{H}_G= \text{span}\{|g_1\rangle, \dots, |g_r\rangle\}$, where $r$ is the ground-state degeneracy and $\{|g_i\rangle\}_{i=1}^{r}$ is an orthonormal basis satisfying $H|g_i\rangle = E_0|g_i\rangle$. Rather than searching for a single state that minimizes $\langle H\rangle$, our task is to generate a collection of output states, whose linear span covers the entire degenerate ground space $\mathcal{H}_G$ efficiently.

Suppose that we obtain a family of states from some methods, denoted by $\{\ket{\psi_m}\}_{m=1}^M$, and collect them as a matrix $C= [ |\psi_1\rangle, \dots, |\psi_M\rangle ] \in \mathbb{C}^{d\times M}$, where $d$ is the physical dimension of the whole system. Besides verifying that each $|\psi_m\rangle$ has energy close to $E_0$, we also quantify the diversity of this family. In particular, our analysis focuses on (i) the number of linearly independent states present in the family $\{\ket{\psi_m}\}_{m=1}^M$, and (ii) the overlap between the span of $\{\ket{\psi_m}\}_{m=1}^M$ and the exact ground space $\mathcal{H}_G$.

To determine the number of linearly independent states in this family, we compute the numerical rank of the matrix $C$, using an error tolerance due to the numerical precision~\cite{FosterLeslieV.2013A9Rc}. Let $\sigma_1(C)\ge \sigma_2(C)\ge \cdots \ge \sigma_{\text{min}(d,M)}(C)$ denote the singular values of $C$. To ensure the rank determination is independent of the overall signal magnitude, we consider the normalized singular values $\tilde{\sigma}_i = \frac{\sigma_i(C)}{\sigma_1(C)}$ ($0 \leq \tilde{\sigma}_i \leq 1 $ for $ 1 \le i \le \min(d, M)$, and $\tilde{\sigma}_1=1$). For a fixed relative error tolerance $\varepsilon>0$, the effective rank is defined as
\begin{equation}
\operatorname{rank}_\varepsilon(C)
= \bigl|\{\, i : \tilde{\sigma}_i > \varepsilon \,\}\bigr|
= \left|\left\{\, i : \frac{\sigma_i(C)}{\sigma_1(C)} > \varepsilon \,\right\}\right|.
\label{eq:tolerancerank}
\end{equation}

The condition $\operatorname{rank}_\varepsilon(C)=r$ indicates that the produced states occupy $r$ linearly independent directions in the degenerate space and therefore accurately infer the dimension of the target ground space. In contrast, if $\operatorname{rank}_\varepsilon(C)<r$, the produced ensemble fails to span the entire space, leaving $r-\operatorname{rank}_\varepsilon(C)$ linearly independent ground states missing.

In addition to the numerical rank, we employ a certification in benchmark cases where an exact ground-state basis is available (e.g., from exact diagonalization). In these settings, we define the ground-space overlap score $p_G(\psi)=\sum_{i=1}^r |\langle g_i|\psi\rangle|$, which we use as a sample-level diagnostic in our benchmark studies. We then compute the principal angles between the span of the produced states and the ground space $\mathcal{H}_G$ to quantify the distance between the generated space and $\mathcal{H}_G$~\cite{kolodrubetzClassifyingMeasuringGeometry2013,Sherif_2026}. Let $U = [|g_1\rangle, \dots, |g_r\rangle] \in \mathbb{C}^{d\times r}$ be a matrix whose orthonormal columns form a basis for $\mathcal{H}_G$, obtained from exact diagonalization. Correspondingly, for the set of states $\{\ket{\psi_m}\}_{m=1}^M$, we construct a matrix $X \in \mathbb{C}^{d\times r}$ such that its columns constitute an orthonormal basis for $\mathrm{span}\{\ket{\psi_1},\dots, \ket{\psi_M}\}$. The principal angles $\alpha_1 , \dots, \alpha_r \in [0,\frac{\pi}{2}]$ between two spaces are defined by
\begin{equation}
\cos \alpha_k = \sigma_k(X^\dagger U), \qquad k \in \{1,\dots,r\}
\label{eq:angles}
\end{equation}
where $X^\dagger U \in \mathbb{C}^{r\times r}$ is the overlap matrix and $\sigma_k(\cdot)$ denotes the corresponding singular value. We use the mean principal angle $\bar{\alpha} = \frac{1}{r}\sum_{k=1}^r \alpha_k,$ the maximal principal angle $\alpha_{\max} = \max_{1\le k \le r} \alpha_k,$ and the chordal distance $d_{\mathrm{ch}}^2 = \sum_{k=1}^r \sin^2 \alpha_k$  to quantify the geometric relationships between the space spanned by$\{\ket{\psi_m}\}_{m=1}^M$ and the true ground space $\mathcal{H}_G$. Small values of $\bar{\alpha}$, $\alpha_{\max}$, and $d_{\mathrm{ch}}^2$---that is, $\bar{\alpha}\approx 0$, $\alpha_{\max}\approx 0$, and $d_{\mathrm{ch}}^2\approx 0$---imply that the space spanned by $\{\ket{\psi_m}\}_{m=1}^M$ has a large overlap with the true ground space $\mathcal{H}_G$, and thus closely approximates the exact ground space.

\section{Architecture and algorithm}

In this section, we present a local-observable-guided generative quantum-circuit framework for learning degenerate ground spaces of quantum many-body Hamiltonians. The method combines a classical generative model with an expressive parameterized quantum circuit (PQC), so that the output is not a single optimized state but an ensemble of quantum states whose span is driven toward the full degenerate ground space. Our method introduces two key ingredients. Firstly, we introduce physically motivated local observables into the training objective, allowing the optimization to effectively distinguish different ground-state directions through accessible one-body and two-body features rather than through full-state information. Secondly, we extend the framework to a finite-shot setting, where both the objective evaluation and gradient estimation must remain stable under sampling noise while keeping the measurement budget manageable.

This section is organized into three parts. We first describe the generative quantum circuit architecture, including the classical generator and the expressive PQC used to prepare quantum states. We then introduce the energy-diversity training objective, which balances energy minimization against a diversity term constructed from local observables and thereby guides the ensemble toward full ground-space coverage. Finally, we present the finite-shot training protocol, including the shot-adaptive schedule and stochastic parameter-shift update used to control measurement cost while maintaining reliable optimization under shot noise.

\subsection{Generative quantum circuit architecture}
\begin{figure}[htbp]
    \centering
    \includegraphics[width=\linewidth]{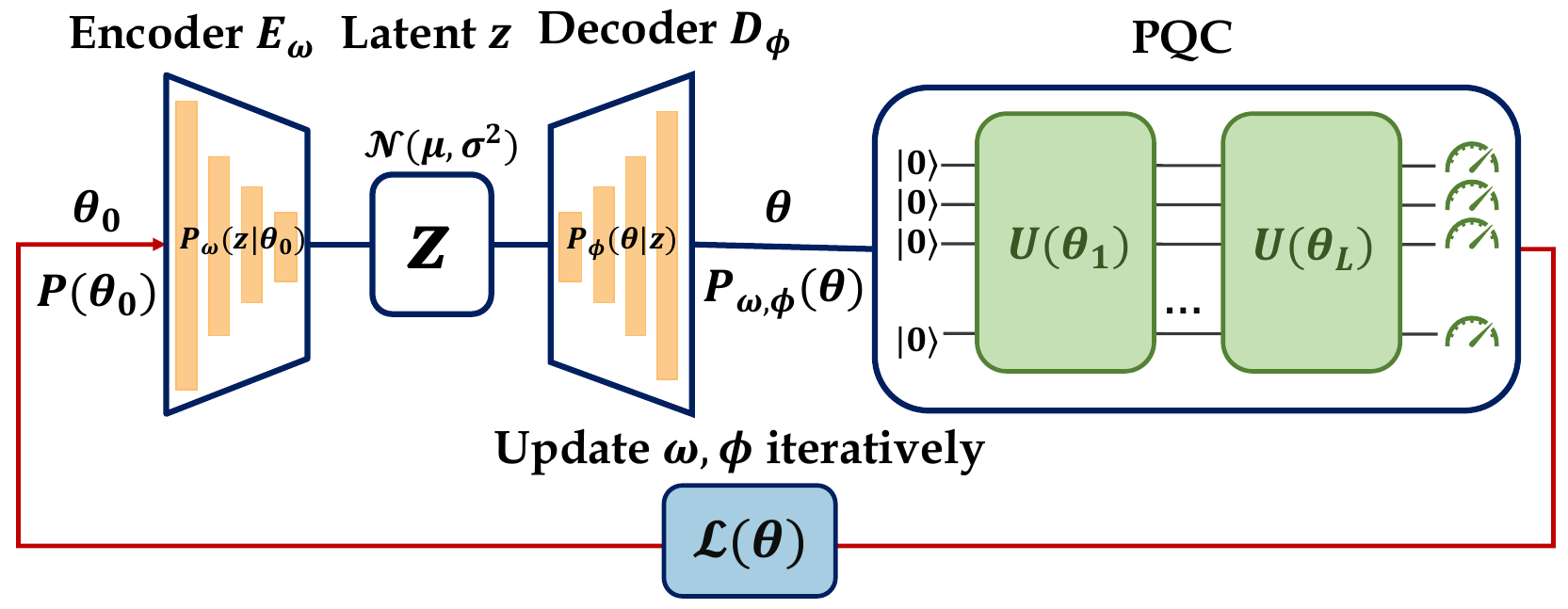}
    \caption{Generative quantum circuit architecture. The network consists of a classical generative model followed by a parametrized quantum circuit (PQC). The encoder $E_{\bm{\omega}}$ maps an input $\bm{\theta}_0 \sim P(\bm{\theta}_0)$ to a low-dimensional latent distribution $\bm{z} \sim \mathcal{N}(\bm{\mu}, \bm{\sigma}^2)$. The decoder $D_{\bm{\phi}}$ then converts $\bm{z}$ into circuit parameters $\bm{\theta} \sim P_{\bm{\omega},\bm{\phi}}(\bm{\theta})$, which configure a PQC with depth $L$. Measurements are then performed to obtain an energy-diversity objective, and the parameters $(\bm{\omega},\bm{\phi})$ are updated iteratively.}
    \label{fig:vgon}
\end{figure}

The network consists of a classical generative model that produces circuit parameters, and a PQC that prepares the corresponding quantum states (see Fig~\ref{fig:vgon}). We refer to this hybrid classical-quantum generative network as a generative quantum circuit. The generative component employs the VGON~\cite{zhangVariationalOptimizationQuantum2025}, which utilizes a pair of deep neural networks in an encoder-decoder configuration, connected by a
stochastic latent layer. The encoder compresses high-dimensional parameters into a low-dimensional latent representation, while the decoder maps these latent variables back to the high-dimensional circuit parameter space. These decoded parameters are then directly fed into the PQC to prepare the corresponding quantum states.

Standard implementations typically employ Rectified Linear Units (ReLU) in the hidden layers and enforce an explicit Kullback-Leibler (KL) regularization penalty in the objective function. However, when applied to degenerate ground states, this conventional configuration often suffers from mode collapse, where the model converges to a single solution instead of capturing the full degenerate ground space. To mitigate this, we replace standard ReLUs with Scaled Exponential Linear Units (SELU)~\cite{klambauerSelfNormalizingNeuralNetworks2017} within the encoder and decoder, and omit the explicit KL penalty. The self-normalizing properties of SELUs naturally induce a stable, approximate Gaussian latent distribution without the rigid constraints of an explicit penalty. By removing the regularization, the model is significantly encouraged to explore the latent space, thereby preserving the diversity essential for generating linearly independent quantum states.

During the training stage, the encoder $E_{\bm{\omega}}$ takes inputs $\bm{\theta}_0 $ sampled from a simple prior $ P(\bm{\theta}_0)$, and outputs parameters $(\bm{\mu}(\bm{\theta}_{0}), \bm{\sigma}(\bm{\theta}_0))$, which define the mean and standard deviation of a low-dimensional Gaussian distribution $P_{\bm{\omega}}(\bm{z})$ over the latent variable $\bm{z} \in \mathbb{R}^{d_{\bm{z}}}$, where $d_{\bm{z}}$ is the dimension of the latent space. More precisely, for each $\bm{\theta}_0$, the encoder defines a conditional distribution $P_{\bm{\omega}}(\bm{z}|\bm{\theta}_0)$, and the latent distribution is given by $P_{\bm{\omega}}(\bm{z}) = \int P_{\bm{\omega}}(\bm{z}|\bm{\theta}_0) P(\bm{\theta}_0) d\bm{\theta}_0 $. The decoder network $D_{\bm{\phi}}$ then defines a conditional distribution $P_{\bm{\phi}}(\bm{\theta}|\bm{z})$ given a latent variable $\bm{z}$, and maps the latent distribution $P_{\bm{\omega}}(\bm{z})$ to a final distribution $P_{\bm{\omega},\bm{\phi}}(\bm{\theta})$ over the circuit parameters $\bm{\theta} \in \mathbb{R}^{d_{\bm{\theta}}}$, where $d_{\bm{\theta}}$ represents the number of variational circuit parameters. Therefore, the final  distribution $P_{\bm{\omega},\bm{\phi}}(\bm{\theta})$ is defined as $P_{\bm{\omega},\bm{\phi}}(\bm{\theta}) = \int P_{\bm{\phi}}(\bm{\theta}|\bm{z})\,P_{\bm{\omega}}(\bm{z})\,d\bm{z}$~\cite{zhangVariationalOptimizationQuantum2025}. Subsequently, the parameters $\bm{\theta}$ sampled from this distribution are fed into the PQC.

A PQC then prepares the state $|\psi(\bm{\theta})\rangle = U(\bm{\theta})\,|0\rangle$. For each batch of generated states $\{|\psi_m(\bm{\theta}_m)\rangle\}_{m=1}^M$, measurements are performed to compute the objective function, which includes the expectation values of both Hamiltonian $ H $ and the local observables that encode distinguishing features of the generated states. Given that the VGON generates a distribution $P_{\bm{\omega},\bm{\phi}}(\bm{\theta})$ over the circuit parameters, we define the loss function as a batch Monte Carlo estimate of the expected objective over sampled parameters, which is expressed as $\mathcal{L}(\bm{\Theta}) = \mathbb{E}_{\bm{\theta}\sim P_{\bm{\bm{\phi},\bm{\omega}}}(\bm{\theta})}\left[\mathcal{L}(\bm{\theta})\right]$, where $\bm{\Theta} = \{\bm{\theta}_1, \dots, \bm{\theta}_M\}$ represents the batch parameters, each $\bm{\theta}_m (m \in \{1, \dots, M\})$ is independently sampled from $P_{\bm{\omega},\bm{\phi}}(\bm{\theta})$. The explicit form of $\mathcal{L}(\bm{\Theta})$ is detailed in Sec.~\ref{sec:objective}.

The classical trainable parameters $(\bm{\omega},\bm{\phi})$ of the encoder and decoder are updated by gradient-based optimization with respect to the loss $\mathcal{L}(\bm{\Theta})$. The training process is terminated once the expected energy of the generated states satisfies a pre-defined convergence criterion. After convergence, the network transitions to the generation stage. In this stage, latent variables are sampled from the learned distribution, decoded into optimal circuit parameters, and fed into the quantum circuit to prepare an ensemble of candidate states with ground-state energy.

\subsection{Energy-diversity training objective}
\label{sec:objective}
Our goal is to generate an ensemble of states that fully spans the degenerate ground space. This requires simultaneously minimizing the system's energy to ensure sample quality and maximizing the distinguishability among generated states to ensure sample diversity. Balancing these requirements involves addressing the quality-diversity trade-off, a fundamental challenge in generative modeling. Specifically, overemphasizing quality typically drives the model toward a limited subset of solutions, whereas enforcing excessive diversity often hinders convergence to the ground states. This intrinsic tension is frequently reflected in transient oscillatory dynamics during the optimization. Consequently, balancing these competing objectives to achieve convergence of the energy to the ground-state value while fully spanning the degenerate ground space remains a non-trivial task in variational quantum algorithms.

A key challenge is defining a diversity objective that is both efficient and physically meaningful. While enforcing diversity by minimizing pairwise state fidelities is conceptually straightforward, it requires significant tomographic resources and does not scale well with system size. In contrast, imposing dispersion directly in gate-parameter space is computationally inexpensive but poorly aligned with the geometry of state space. Distances in parameter space do not faithfully reflect distances between quantum states, so parameter-based separation of degenerate ground states is unreliable in many models.

A physically motivated route is suggested by the local-data reconstruction results~\cite{swingle2014reconstructing,huangPredictingManyProperties2020a,cramerEfficientQuantumState2010a}. These works show that for a broad class of local Hamiltonians, ground states can be reconstructed from local measurements and classical postprocessing. A particularly transparent case is provided by frustration-free local Hamiltonians, which can be written as $H = \sum_i h_i$. In such models, the global ground space of $H$ is also in the local ground spaces of $h_i$, so that the essential structure of the degenerate ground space is encoded in reduced density matrices~\cite{chen2012ground}. This perspective motivates enforcing diversity directly in the space of local observables, rather than in fidelity space or in the space of circuit parameters.

Guided by these insights, we construct the diversity objective by projecting each generated quantum state $\ket{\psi(\bm{\theta})}$ onto a feature space defined by a selected set of observables $\mathcal{O} = \{ \hat{O}_1, \dots, \hat{O}_D \}$. Here, the operators $\hat{O}_j (j \in \{1, \dots, D\})$ are typically chosen as Pauli strings that characterize the distinguishing correlations of the ground states. This mapping yields a real-valued feature vector $\bm{l}(\bm{\theta})$ composed of the expectation values:
\begin{equation}
\bm{l}(\bm{\theta}) = \big( \langle \hat{O}_1 \rangle, \dots, \langle \hat{O}_D \rangle \big) \in \mathbb{R}^{D}
\end{equation}

For a broad class of physical models, such as frustration-free Hamiltonians, the degenerate ground states can be effectively distinguished via their local one- and two-body correlations. Consequently, we define the feature vectors using the expectation values of one-body and nearest-neighbor two-body Pauli operators:
\begin{equation}
\begin{split}
\bm{l}^{(1)}(\bm{\theta})
&=\big( \langle \sigma_1^\alpha\rangle, \dots, \langle \sigma_N^\alpha\rangle \big)
\ \in\ \mathbb{R}^{3N} \\
\bm{l}^{(2)}(\bm{\theta})
&=\Big( \langle \sigma_1^\alpha \sigma_2^\beta\rangle, \dots,  \langle \sigma_N^\alpha \sigma_1^\beta\rangle  \Big)
\ \in\ \mathbb{R}^{9N}
\end{split}
\label{eq:one/twocorre}
\end{equation}
where $\sigma_i^\alpha$ is the Pauli matrix with $\alpha\in\{x,y,z\}$ acting on site $i (i \in \{1, \dots, N\})$. If additional prior information about the locally distinguishable features of ground states is available, the measured sets can be replaced accordingly to reduce measurement cost. We note that the framework is extensible to systems possessing long-range entanglement. In such cases, the feature set can be replaced by suitable non-local operators to capture the relevant global information~\cite{li2022detecting}.

We quantify the diversity among the generated states using the cosine similarity of these feature vectors within a batch. For a batch of generated states $\{|\psi_m(\bm{\theta}_m)\rangle\}_{m=1}^M$, the diversity loss $\mathcal{L}^{(k)}$ is defined as the batch-mean cosine similarity between their corresponding feature vectors:
\begin{equation}
\mathcal{L}^{(k)}(\bm{\Theta})=\frac{2}{M(M-1)}\sum_{1\le i<j\le M}\frac{\langle \bm{l}^{(k)}(\bm{\theta}_i), \bm{l}^{(k)}(\bm{\theta}_j)\rangle}{\|\bm{l}^{(k)}(\bm{\theta}_i)\| \|\bm{l}^{(k)}(\bm{\theta}_j)\|}
\label{eq:cosine}
\end{equation}
where $\mathcal{L}^{(k)} \in [-1,1]$, and $k\in\{1,2\}$ corresponds to the one- and two-body feature sets, respectively. $\mathcal{L}^{(k)} \approx 1$ signifies that the generated states share identical features, corresponding to mode collapse, $\mathcal{L}^{(k)} \approx 0$ indicates orthogonal feature vectors, and $\mathcal{L}^{(k)} \approx -1$ denotes opposing correlations. Minimizing $\mathcal{L}^{(k)}$ encourages the feature vectors to spread out in the feature space, thereby driving the generated states to exhibit distinct physical correlations.

Combining the energy minimization requirement with the diversity term defined above, the total energy-diversity objective function is given by
\begin{equation}
\begin{split}
    \mathcal{L}(\bm{\Theta})=&
\frac{1}{M} \sum_{m=1}^M \lambda_1 \bra{\psi_m(\bm{\theta}_m)}H\ket{\psi_m(\bm{\theta}_m)}
\\
&+\lambda_2 (\mathcal{L}^{(1)}(\bm{\Theta})+ \mathcal{L}^{(2)}(\bm{\Theta}))
\label{eq:loss}
\end{split}
\end{equation}
where $\lambda_1,\lambda_2 >0$ are the corresponding coefficients that balance the energy and diversity contributions. Lower values of $\mathcal{L}^{(1)}(\bm{\Theta})$ and $\mathcal{L}^{(2)}(\bm{\Theta})$ correspond to greater dispersion in the local-observable features, ensuring the model explores a diverse set of states across the degenerate space.

\subsection{Training under finite-shot noise}
\label{sec:finite-shot}
In the generative quantum-circuit architecture, the training objective is built from the Hamiltonian energy and selected local observables evaluated over a batch of generated circuits. In an ideal numerical simulation, these quantities are available exactly. In the finite-shot setting, however, they are estimated from a finite number of measurement samples, so sampling noise enters both the loss and the optimization. This is particularly challenging in the present multiobjective setting, where one must simultaneously drive the ensemble toward the ground space and preserve sufficient diversity to recover the full degenerate ground space. Finite-shot fluctuations can therefore perturb the balance between the energy and diversity terms and increase the risk of mode collapse. A second practical challenge is gradient evaluation. Under shot-based estimation, standard approaches such as parameter-shift differentiation require additional circuit executions for each trainable parameter, so the measurement cost grows rapidly with circuit size. Reliable finite-shot training must therefore address both the destabilizing effect of sampling noise on the multiobjective loss and the overhead of gradient estimation within a practical measurement budget.

Despite these challenges, the generative quantum circuit admits a natural shot-adaptive training strategy. The key point is that each update is determined by batch-level statistics rather than by the precise evaluation of any single circuit. Batch-averaged energies and batch-level cosine similarities partially average out finite-shot noise across the ensemble, making the optimization less sensitive to fluctuations in individual samples. As a result, in the early stage of training it is sufficient to resolve only coarse distinctions between higher- and lower-energy regions while maintaining dispersion across the batch, so a small shot budget is already effective. We therefore begin with a low shot count and increase it only after the sampled batch approaches the ground space, where more accurate energy and feature estimates are needed to stabilize the update and ensure reliable convergence.

To control the measurement cost of gradient evaluation, we use a stochastic parameter-shift update~\cite{Banchi2021jmc}. As the circuit depth increases, the number of trainable parameters grows, and a full parameter-shift gradient quickly becomes prohibitively expensive, which requires two shifted circuit evaluations per parameter. At each iteration, we therefore randomly choose a subset $\mathcal{I}_K \subset \{1,\dots,n_p\}$ of $K$ active parameters, where $n_p$ is the total number of trainable parameters in the circuit and $K\ll n_p$. For each sample in the batch, with circuit parameters $\bm{\theta}_m \in \mathbb{R}^{n_p}$ $(m=1,\dots,M)$, we estimate the loss gradient only along these active coordinates using the parameter-shift rule with a fixed shift $\delta=\pi/2$:
\begin{equation}
    \partial_{\theta_{m,j}} \mathcal{L}
\approx
\frac{1}{2}
\Big[
\mathcal{L}\big(\bm{\theta}_m+\delta\,\bm e_j\big)
-
\mathcal{L}\big(\bm{\theta}_m-\delta\,\bm e_j\big)
\Big],
\qquad j\in\mathcal{I}_K
\end{equation}
where $\theta_{m,j}$ denotes the $j$th component of $\bm{\theta}_m$, $\bm e_j$ is the $j$th unit vector in $\mathbb{R}^{n_p}$, and $\mathcal{L}$ is the finite-shot loss in Eq.~(\ref{eq:loss}). With this choice, each iteration requires only $2K$ shifted circuit evaluations for each batch sample, significantly reducing the measurement overhead.

The finite-shot protocol is benchmarked on the MG chain at small system sizes ($N=5,6$), for which shot-based training and certification remain computationally feasible. The corresponding results are reported in Sec.~\ref{sec:mg_shots}.

\section{Results on many-body models}
\label{sec:cases}
The key observation behind our approach is that, in a broad class of systems with ground-state degeneracy, different ground states exhibit distinguishable features at the level of one- and two-body correlators. To demonstrate that the generative quantum circuit and energy-diversity training objective are applicable to a broad class of models, we consider three representative models: (i) the MG model, where degeneracy arises from frustration, (ii) the AKLT model, a paradigmatic example of symmetry-protected topological order, and (iii) the spin-1 XXZ chain, where the degeneracy is associated with the first-order phase transition at $\Delta = -1$.

\begin{table*}[htbp]
  \centering
  \caption{ Training configuration for each model, including the number of sites $N$ in encoded quantum circuit, the encoder structure $E$, decoder structure $D$, latent space dimension $d_z$, circuit depth $L$, number of circuit parameters $n_p$, mini-batch size $M$, and learning rate $\eta $. }
  \begin{tabular}{lccccccccc}
  \toprule
  &$\text{Model}$ &$N$ &$E$ &$D$ &$d_z$  &$L$ &$n_p$ &$M$ &$\eta (\times 10^{-4})$ \\
  \midrule
   &$\text{MG}$ &$9$  &$[512,256,128,64]$ &$[64,128,256,512]$ &$50 $ &$5$  &$600$ &$50$ &$3$\\
   &$\text{MG}$ &$10$ &$[512,256,128,64]$  &$[64,128,256,512]$ & $50 $ & $6$    &$810$ &$70$ &$3$\\
   &$\text{AKLT}$ &$10$   &$[512,256,128,64,32]$  & $[32,64,128,256,512]$  & $30 $ & $5$  &$285$  &$50$ &$4$\\
   &$\text{spin-1 XXZ}$ &$8$  &$[512,256,128,64,32,16]$  & $[16,32,64,128,256,512]$  &$50 $ &$6$   &$270$ &$100$ &$6$\\
  \bottomrule
  \end{tabular}
  \label{tab:configuration}
\end{table*}



\subsection{The Majumdar--Ghosh (MG) model}
The MG model is a spin-$\tfrac{1}{2}$ chain with nearest- and next-nearest-neighbor interactions, taken under open boundary conditions (OBC) as
\begin{equation}
H_{\mathrm{MG}}=\sum_{i=1}^{N-2} (\mathbf  S_i \cdot \mathbf S_{i+1} + \mathbf S_{i+1} \cdot \mathbf S_{i+2} + \mathbf S_i \cdot \mathbf S_{i+2} )
\label{eq:MG}
\end{equation}
where $\mathbf S_i$ are spin-$\tfrac{1}{2}$ operators at site $i$. The competition between nearest-neighbor and next-nearest-neighbor interactions drives the MG chain into a frustrated, spontaneously dimerized phase with a nontrivial ground-state structure~\cite{zhangVariationalOptimizationQuantum2025,caspersMajumdarGhoshChainTwofold1984,chhajlanyEntanglementMajumdarGhoshModel2007}. The resulting ground space is spanned by valence-bond coverings that pair spins into singlets on alternating bonds (even versus odd). Under OBC, boundary constraints lead to a size-dependent degeneracy, four-fold for odd $N$ and five-fold for even $N$~\cite{zhangVariationalOptimizationQuantum2025}.

\subsubsection{Energy-diversity objective and training dynamics}

We benchmark the proposed generative quantum circuit on the Majumdar-Ghosh (MG) chain with OBC at system sizes $N=9$ and $N=10$, whose exact ground-state energies are $E_0=-21$ and $E_0=-24$, respectively. For both system sizes, the classical generator uses encoder and decoder networks with four hidden layers, and the latent dimension is fixed at $d_z=50$. The circuit depth is chosen as $L=5$ for $N=9$ and $L=6$ for $N=10$. Training is performed with the Adam optimizer at an initial learning rate $\eta=3\times 10^{-4}$, which is gradually decayed during optimization. The full configuration is summarized in Table~\ref{tab:configuration}, and the PQC employs the expressive circuit ansatz described in Appendix~\ref{app:mg-ansatze}.

\begin{figure}[htbp]
	\centering
	\begin{subfigure}[t]{0.3\textwidth}
		\centering
		\includegraphics[width=\linewidth]{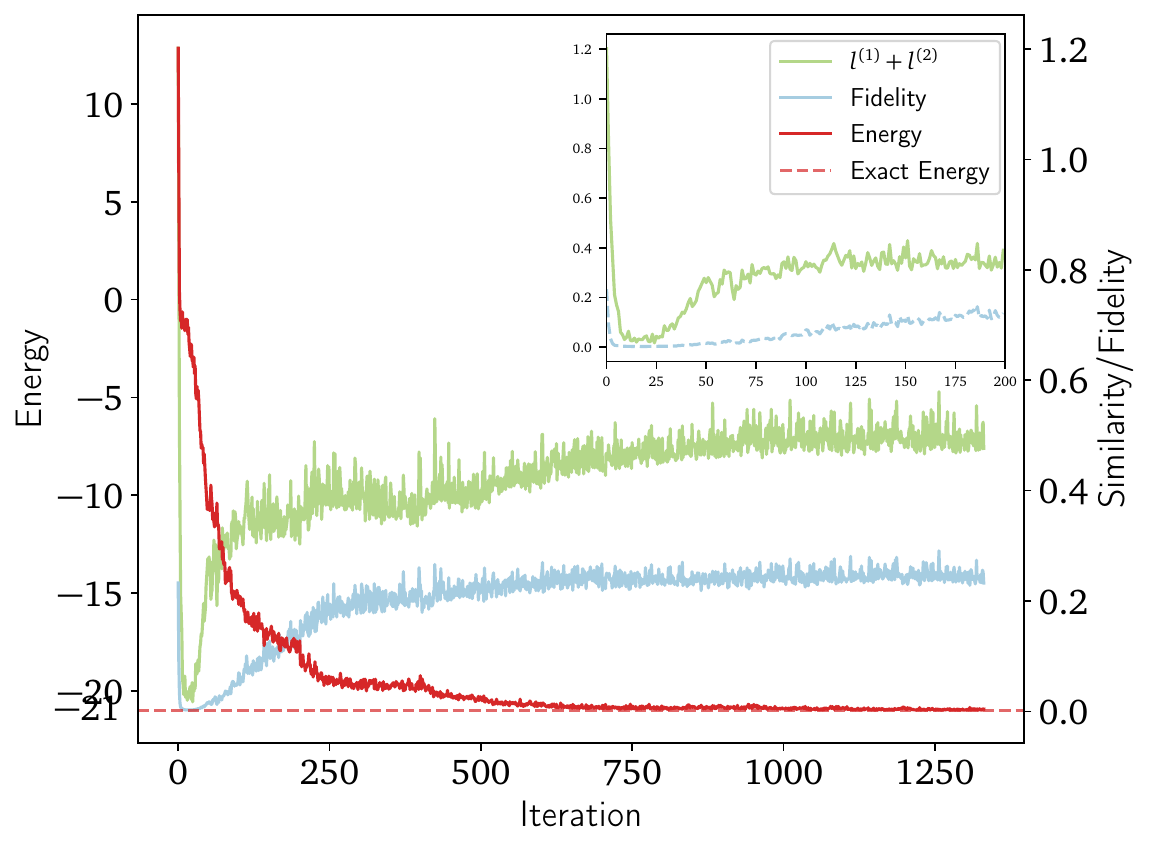}
		\caption{system size $N=9$} 
		\label{fig:mg-dyn-N9}
	\end{subfigure}
	\hspace{0.1mm}
	\begin{subfigure}[t]{0.3\textwidth}
		\centering
		\includegraphics[width=\linewidth]{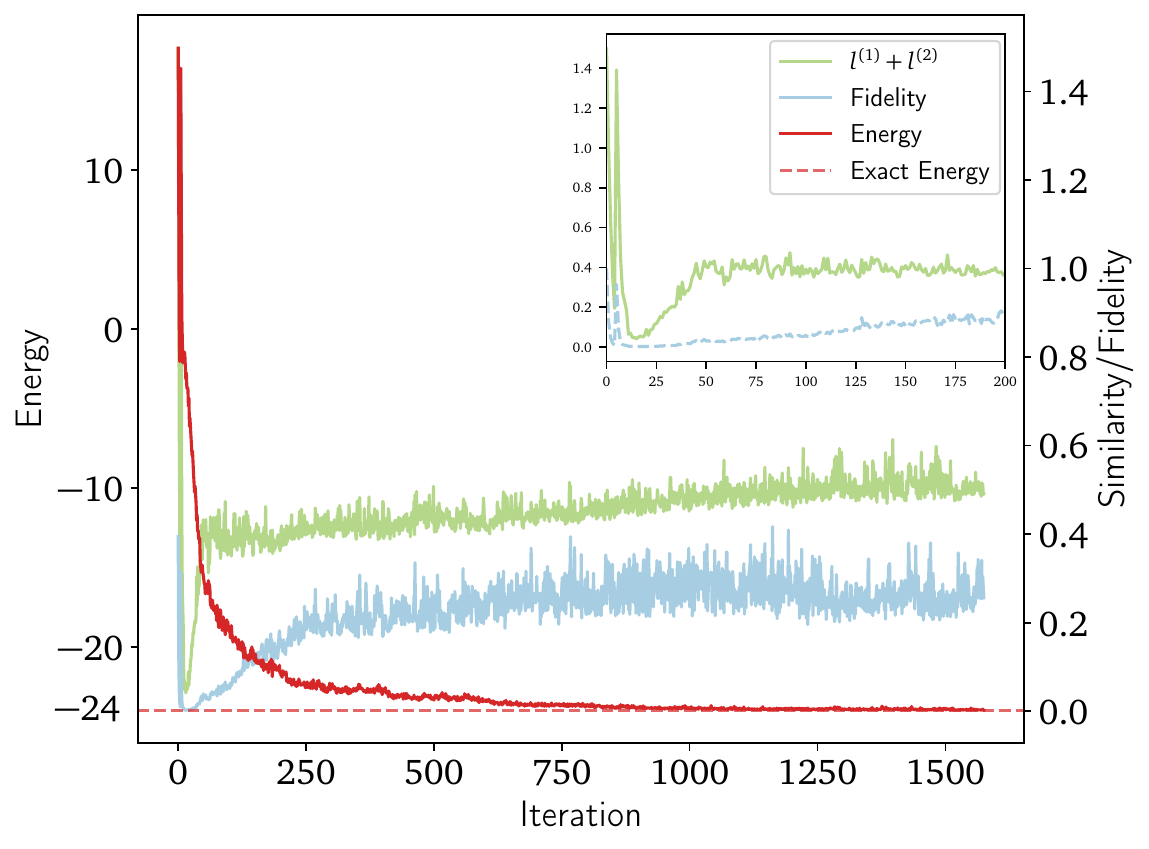}
		\caption{system size $N=10$} 
		\label{fig:mg-dyn-N10}
	\end{subfigure}
\caption{Training dynamics for the MG model. Shown are the batch-mean energy (red), the batch-mean pairwise fidelity (blue), and the sum of cosine similarities $\bm{l}^{(1)} + \bm{l}^{(2)}$ (green). The insets highlight the first 200 iterations. The dynamics exhibit an initial diversity-dominated spreading stage followed by an energy-dominated concentration into the degenerate ground space. The persistent fluctuations reflect the competition between energy minimization and diversity preservation.}
	\label{fig:mg-dynamics}
\end{figure}

For each batch, the objective combines the Hamiltonian expectation with cosine-similarity penalties built from the local feature vectors in Eq.~\eqref{eq:loss}. Here the one-body vector $\bm{l}^{(1)}$ encodes local magnetizations, while the nearest-neighbor two-body vector $\bm{l}^{(2)}$ captures correlations sensitive to the underlying dimer pattern. These quantities allow us to monitor how the generated ensemble evolves in both energy and local-observable feature space during training.

The resulting dynamics, shown in Fig.~\ref{fig:mg-dynamics}, reveal a clear two-stage training process. We track three batch-averaged quantities: the combined cosine similarity $\bm{l}^{(1)} + \bm{l}^{(2)} \in [-2,2]$, the pairwise fidelity $ |\langle \psi_i|\psi_j\rangle|^2 $ between distinct generated states in the batch, and the batch-mean energy. At the beginning of training, the diversity weight is set to a large value, $\lambda_2=10$, while the energy weight is fixed at $\lambda_1=1$, so the optimization is initially dominated by the diversity term. Correspondingly, both the batch-mean cosine similarity and the batch-mean pairwise fidelity drop rapidly, indicating that the generated states spread across Hilbert space rather than collapsing onto a single configuration. In this regime, the objective promotes broad exploration of the accessible space. 

As training proceeds, $\lambda_2$ is gradually annealed to $0.2$, and the energy term becomes the dominant contribution. The batch-mean energy is then driven toward the exact ground-state value, while the cosine similarity and pairwise fidelity increase from their early-time minima, indicating that the ensemble is concentrating within the ground space rather than continuing to spread across Hilbert space. Importantly, both quantities remain well below 1, showing that the ensemble does not collapse onto a single state. Instead, the generated states concentrate into the degenerate ground space while retaining internal diversity. The residual fluctuations in all three quantities reflect the persistent competition between lowering the energy and maintaining ensemble diversity. Taken together, these dynamics show that the energy-diversity objective first drives efficient exploration and then concentrates the ensemble into the degenerate ground space.

To determine when this process has converged, we impose a termination criterion based on both the batch-mean energy and the worst sample in the batch. Specifically, training is halted only when $|E-E_0|<0.02$ and $|E_{\max}-E_0|<0.08$ have been satisfied in 5 iterations in total. This condition ensures not only that the mean energy is close to the exact ground-state value, but also that the entire batch is uniformly concentrated near the target space.

\subsubsection{Characterization of the learned ground space}
After satisfying the convergence criteria, we terminate training and assess the learned ensemble by sampling $1500$ latent variables for each system size. A generated state is accepted if it simultaneously satisfies a relative energy constraint $\Delta E=\frac{|E-E_0|}{|E_0|}<0.005$ and a ground-space overlap score $p_G(\psi)=\sum_{i=1}^r |\langle g_i|\psi\rangle|>0.995$. Applying these criteria, we obtain acceptance rates of $90.9\%$ for $N=9$ ($E_0=-21$) and $90.3\%$ for $N=10$ ($E_0=-24$). These high acceptance rates show that the trained generative quantum circuit reliably produces ground states within the target space.
\begin{table}[htbp]
  \centering
  \caption{Metrics characterizing the recovered ground space for each model encoded to $N$-sites quantum circuit, including the tolerance rank $\text{rank}_{\epsilon}(C)$, mean and maximum principal angles $(\bar\alpha, \alpha_{\max})$, and chordal distance $d_{\mathrm{ch}}^2$.}
  \begin{tabular}{lcccccc}
  \toprule
  &$\text{Model}$ &$N$  &$\text{rank}_{\epsilon}(C)$ &$\bar\alpha (\times 10^{-6})$ & $\alpha_{\max}(\times 10^{-6})$ & $d_{\mathrm{ch}}^2 (\times 10^{-15})$ \\
  \midrule
   &\text{MG} &9    &4 & $0.942 $ & $1.71 $ & $1.78 $ \\
   &\text{MG} &10   &5 & $1.21$ & $2.415$ & $3.55$ \\
   &\text{AKLT} &10  &4 & $2.16$ & $2.7$ & $6.66$ \\
   &\text{spin-1 XXZ} &8  &9 & $0.259$ & $1.48$ & $0$ \\
  \bottomrule
  \end{tabular}
  \label{tab:metrics}
\end{table}

We then assess whether the accepted states cover the full degenerate ground space. To this end, we assemble the accepted states into the coefficient matrix $C$ defined in Sec.~\ref{sec:Problem} and evaluate its numerical rank with tolerance $\varepsilon=0.05$. In both system sizes, the tolerance rank reproduces the exact degeneracy, namely $\operatorname{rank}_{\varepsilon}(C)=4$ for $N=9$ and $\operatorname{rank}_{\varepsilon}(C)=5$ for $N=10$. The corresponding singular values are $\{1,0.9539,0.9152,0.8733\}$ for $N=9$ and $\{1,0.8373,0.7887,0.7438,0.6337\}$ for $N=10$. All values are significantly larger than the tolerance threshold $\varepsilon$, which shows that the accepted ensemble spans the full degenerate ground space with clear numerical stability.
\begin{figure}[htbp]
  \centering
  \begin{subfigure}[t]{0.235\textwidth}
    \centering
    \includegraphics[width=\linewidth]{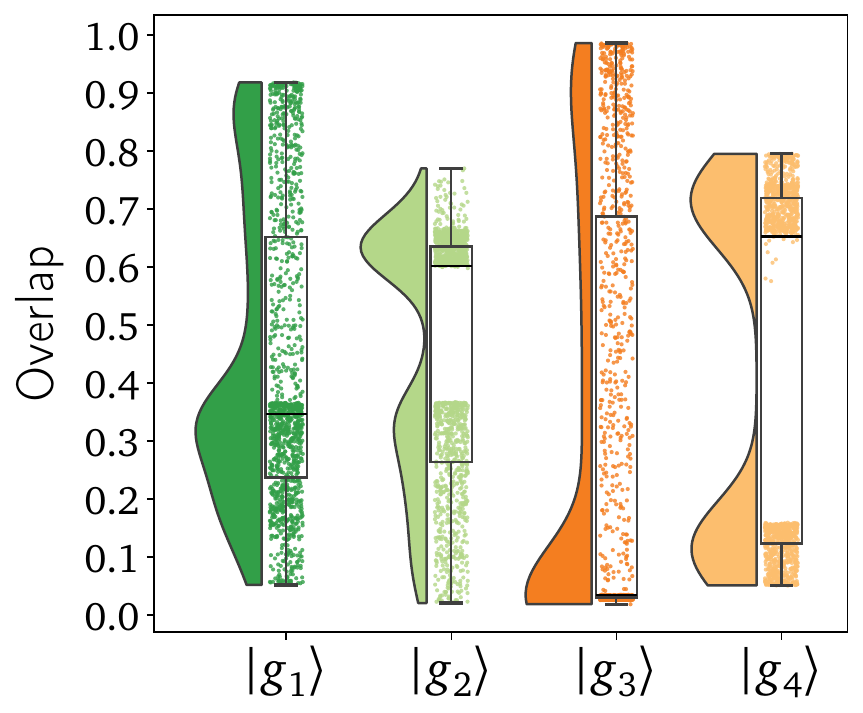}
    \caption{ system size $N=9$ with degeneracy $k=4$}
  \end{subfigure}
  \hspace{0.05mm}
  \begin{subfigure}[t]{0.235\textwidth}
    \centering
    \includegraphics[width=\linewidth]{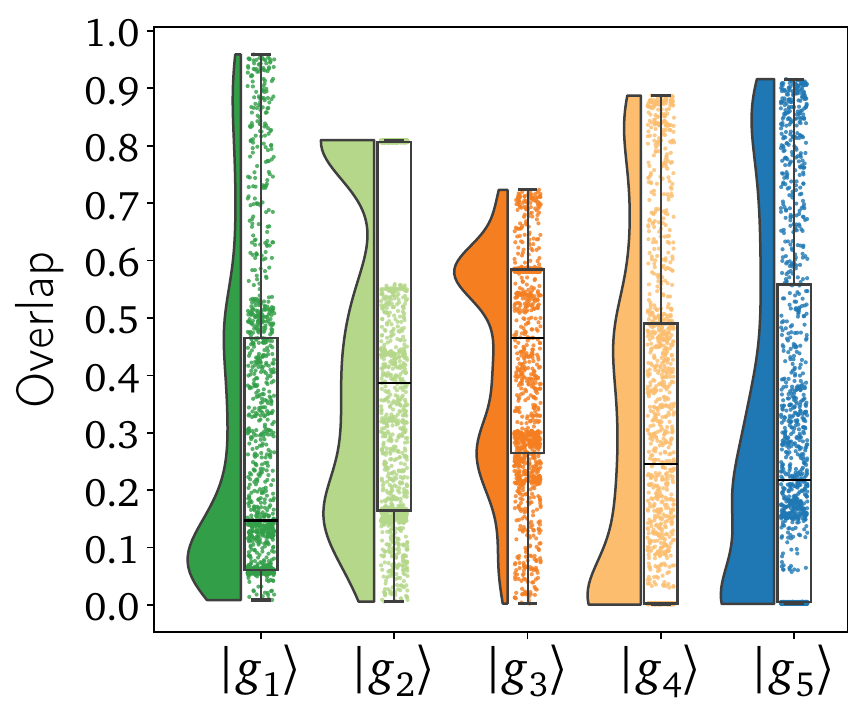}
    \caption{system size $N=10$ with degeneracy $k=5$}
  \end{subfigure}
  \caption{Overlap distributions of accepted states $\{|\psi_i\rangle\}_{i=1}^{1500}$ with the MG ground-state basis $\{|g_i\rangle\}_{i=1}^{k}$. The plots combine probability densities (the half-violin plot on the left) with quartile statistics (the box-and-whisker plot on the right box). The broad distributions indicate a diverse generated ensemble covering the full degenerate ground space.}
  \label{fig:mg-overlaps}
\end{figure}

To visualize the diversity of the accepted states within the generated ensemble, Fig.~\ref{fig:mg-overlaps} presents the overlap distributions of the accepted states with respect to the orthonormal ground-state basis. For both $N=9$ and $N=10$, each scatter point represents the overlap of a single generated sample $|\psi_i \rangle$, with its magnitude $|\langle g_j | \psi_i \rangle|$ onto the $j$-th ground-state basis. The generated ensemble therefore does not concentrate on only a few representative states. Rather, the broad overlap distributions across all basis states indicate that the learned ensemble explores all different directions of the degenerate ground space in a relatively balanced way.

The geometric relation between the learned span and the exact ground space $\mathcal{H}_G$ is further quantified by the principal angles and the chordal distance reported in Table~\ref{tab:metrics}. For the MG model, both the mean and maximum principal angles, $(\bar\alpha, \alpha_{\max})$, are extremely small, and the corresponding chordal distances $d_{\mathrm{ch}}^2$ are close to zero. Together with the full tolerance rank and the high acceptance rates, these metrics show that the space generated by our method coincides with $\mathcal{H}_G$ to numerical accuracy. 

Beyond this space-level agreement, we further find that the generated ensemble contains a set of states forming an approximately orthogonal basis of the ground space (see App.~\ref{app:mg}). The cosine-similarity structure of these states closely matches that of the exact ground-state basis, showing that the generative quantum circuit captures not only the correct ground-space span but also a meaningful basis structure within feature space.

\subsubsection{Robustness to finite-shot sampling}
\label{sec:mg_shots}
We further assess the robustness of the generative quantum circuits in a finite-shot setting. This benchmark is carried out on the MG chain at small system sizes $N=5,6$, with expectation-value simulations at the same sizes serving as a noiseless reference. The finite-shot and expectation-value studies use the same circuit architecture and optimization hyperparameters (Table~\ref{tab:configuration-shots}), and differ only in how observables and gradients are evaluated, as detailed in Sec.~\ref{sec:finite-shot}. For the finite-shot setting, the stochastic parameter-shift update uses $K=50$ active parameters for $N=5$ and $K=80$ for $N=6$, and the shot number is increased during training from $10^3$ to $8\times10^3$, and finally to $1.2\times10^4$ near convergence.
\begin{table}[htbp]
  \centering
  \caption{Training configuration for the MG model, including the number of sites $N$ encoded in the quantum circuit, the encoder structure $E$, decoder structure $D$, latent space dimension $d_z$, circuit depth $L$, number of circuit parameters $n_p$, mini-batch size $M$, learning rate $\eta$, and number of active circuit parameters $K$.}
  \begin{tabular}{ccccccccc}
  \toprule
  $N$ & $E$ & $D$ & $d_z$ & $L$ & $n_p$ & $M$ & $\eta (\times 10^{-4})$ &$K$ \\
  \midrule
  $5$ & $[128,64,32]$ & $[32,64,128]$ & $20$ & $3$ & $180$ & $50$  & $3$ &$50$ \\
  $6$ & $[128,64,32]$ & $[32,64,128]$ & $30$ & $3$ & $225$ & $100$ & $3$ &$80$\\
  \bottomrule
  \end{tabular}
  \label{tab:configuration-shots}
\end{table}

After convergence, we form an ensemble of states by sampling latent variables, decoding them into circuit parameters, and executing the corresponding circuits. For both system sizes, we apply the same certification criteria as in the expectation-value study. A state is accepted if it satisfies the relative energy constraint $\Delta E=\frac{|E-E_0|}{|E_0|}<0.01$ with $E_0=-9$ for $N=5$ and $E_0=-12$ for $N=6$, together with the ground-space overlap weight $p_G(\psi)=\sum_{i=1}^{r}|\langle g_i|\psi\rangle|^2>0.99.$ Using $1500$ latent samples for each system size, the finite-shot acceptance rate is $78.6\%$ for $N=5$ and $70.4\%$ for $N=6$, compared with $90.1\%$ and $90.5\%$ in the expectation-value reference. This reduction is consistent with finite-sampling error in the shot-based energy and feature estimates, but the acceptance rates remain sufficiently high to support reliable recovery of the degenerate ground space.
\begin{table}[htbp]
  \centering
  \caption{Span metrics for the MG ground space under expectation-value and finite-shot evaluation. Reported are the tolerance rank $\mathrm{rank}_{\varepsilon}(C)$, the mean and maximum principal angles $(\bar\alpha,\alpha_{\max})$, and the chordal distance $d_{\mathrm{ch}}^2$.}
  \begin{tabular}{l c c c c c}
  \toprule
  $\text{Evaluation}$ & $N$ & $\mathrm{rank}_{\varepsilon}(C)$ & $\bar\alpha (\times 10^{-6})$ & $\alpha_{\max}(\times 10^{-6})$ & $d_{\mathrm{ch}}^2 (\times 10^{-15})$ \\
  \midrule
  Finite-shot  & 5 & 4 & $1.91$ & $1.21$  & $2.66$ \\
  Expval       & 5 & 4 & $1.47$ & $0.817$ & $1.33$ \\
  Finite-shot  & 6 & 5 & $2.09$ & $1.40$  & $3.55$ \\
  Expval       & 6 & 5 & $1.21$ & $0.428$ & $0.88$ \\
  \bottomrule
  \end{tabular}
  \label{tab:metrics-shots}
\end{table}

The more important question is whether this reduction in acceptance affects the recovered ground space. To address this, we evaluate the span metrics of the accepted ensemble. As summarized in Table~\ref{tab:metrics-shots}, the numerical rank with tolerance $\varepsilon=0.05$ reproduces the exact degeneracy in both settings, with $\mathrm{rank}_{\varepsilon}(C)=4$ for $N=5$ and $\mathrm{rank}_{\varepsilon}(C)=5$ for $N=6$. The principal angles and chordal distances also remain small, indicating that the learned span under finite-shot estimation is closely aligned with the target ground space $\mathcal{H}_G$. Taken together, these results show that finite-shot noise primarily lowers the certification yield, while leaving the recovered ground-space coverage essentially unchanged.

\subsection{The Affleck--Kennedy--Lieb--Tasaki (AKLT) model}

The AKLT model is a spin-1 chain with bilinear-biquadratic interactions under OBC, defined as 
\begin{equation}
H_{\mathrm{AKLT}}
=\sum_{i=1}^{N-1} \left(\mathbf S_i \cdot \mathbf S_{i+1}
+\tfrac{1}{3} (\mathbf S_i \cdot \mathbf S_{i+1})^2\right),
\label{eq:AKLT}
\end{equation}
where $\mathbf S_i$ are spin-1 operators on site $i$. Its ground states form the valence-bond solid (VBS) space. Each spin-1 is represented by two virtual spin-$\tfrac12$ that form singlets on nearest-neighbor links. Under OBC, one virtual spin-$\tfrac12$ remains at each boundary, producing a fourfold ground-space degeneracy carried by edge modes~\cite{affleck1987valence,affleck1987rigorous}.

\subsubsection{Energy-diversity objective and training dynamics}

To implement the AKLT model on variational qubit circuits, we adopt the symmetry-encoding scheme introduced in Ref.~\cite{yangCostLocallyApproximating2025}, in which each spin-1 degree of freedom is embedded into a two-qubit symmetric subspace by a fixed local isometry. The variational circuit is then restricted to symmetry-preserving layers acting within the encoded space. Rather than using arbitrary two-qutrit unitaries, we employ a general ansatz with a reduced number of parameters, which retains sufficient expressiveness while simplifying the circuit design. Full details of the encoding and circuit construction are given in Appendix~\ref{app:spin-ansatze}. In the present case, we implement an $N'=5$ qutrit AKLT chain through a qubit circuit of size $N=10$, whose exact ground-state energy is $E_0=-2.5298$.
\begin{figure}[htbp]
	\centering
	\includegraphics[width=0.65\linewidth]{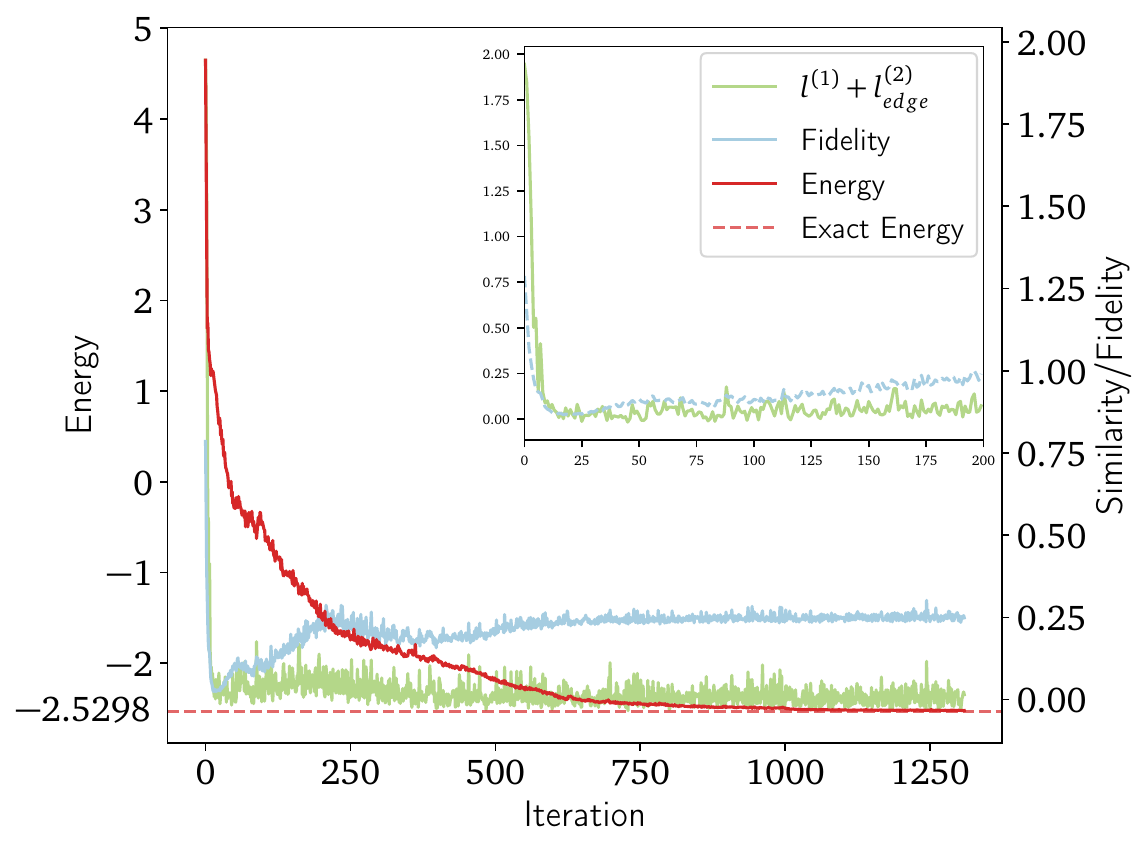}
	\caption{Training dynamics for the symmetry-encoded AKLT model at system size $N=10$. Shown are the batch-mean energy (red), the batch-mean pairwise fidelity (blue), and the sum of cosine similarities  $\bm{l}^{(1)} + \bm{l}^{(2)}_{\mathrm{edge}}$ (green). The inset highlights the first 200 iterations. The dynamics exhibit an initial diversity-dominated spreading stage followed by an energy-dominated concentration into the degenerate ground space. Finally, the energy converges to the ground-state energy with low batch-mean pairwise fidelity.}
	\label{fig:aklt-dynamics}
\end{figure}

For this model, the classical generator uses a five-layer encoder and decoder, with latent dimension $d_z=30$, and the variational circuit has depth $L=5$. Training is performed with the Adam optimizer, starting from a learning rate $\eta=4\times 10^{-4}$ that is gradually decayed during optimization. The full set of hyperparameters is summarized in Table~\ref{tab:configuration}. Considering the edge-localized structure of the AKLT ground space, we define the diversity term by combining the one-body feature vector $\bm{l}^{(1)}$ in Eq.~(\ref{eq:one/twocorre}) with an edge-restricted two-body feature vector
\begin{equation}
    \bm{l}_{\mathrm{edge}}^{(2)}(\bm{\theta})=
    \bigl(
    \langle \sigma_{1}^{\alpha}\sigma_{2}^{\beta}\rangle,\,
    \langle \sigma_{N-1}^{\alpha}\sigma_{N}^{\beta}\rangle,\,
    \langle \sigma_{N}^{\alpha}\sigma_{1}^{\beta}\rangle
  \bigr)\in \mathbb{R}^{27},
  \label{eq:aklt-two-edge}
\end{equation}
where $\alpha,\beta\in\{x,y,z\}$. These observables provide a compact description of the edge degrees of freedom that distinguish the degenerate AKLT ground states.

For the symmetry-encoded AKLT model, the training dynamics in Fig.~\ref{fig:aklt-dynamics} show that our energy-diversity strategy remains robust, successfully guiding the circuit to cover the degenerate ground space. As in the MG case, we monitor three batch-averaged quantities throughout training: the combined cosine similarity built from $\bm{l}^{(1)}+\bm{l}^{(2)}_{\mathrm{edge}} \in [-2,2]$, the pairwise fidelity between distinct generated states in the batch, and the batch-mean energy. The overall behavior of the key quantities in training dynamics closely mirrors the behavior observed in the MG model. At the beginning of training, the diversity weight is set to $\lambda_2=4$, so the optimization first emphasizes the spreading of generated states. In this regime, both the combined cosine similarity and the batch-mean fidelity drop rapidly, as highlighted in the inset of Fig.~\ref{fig:aklt-dynamics}, indicating that the generated states quickly separate in Hilbert space rather than collapsing onto a single configuration.

As training proceeds, $\lambda_2$ is gradually annealed to $0.1$, and the energy term becomes dominant. The batch-mean energy is then driven toward the exact ground-state value, while the similarity and fidelity rise, indicating that the ensemble is no longer spreading freely but is instead becoming concentrated within the ground space. Importantly, however, these quantities remain low, showing that the generated states do not collapse to a single state even as the energy converges. The slight oscillations in the late stage reflect the persistent competition between lowering the energy and preserving diversity among the generated samples. Taken together, these dynamics show that the energy-diversity objective remains effective in the symmetry-encoded AKLT setting. It first promotes broad exploration of the relevant state space and then concentrates the ensemble into the degenerate ground space.
 
To determine convergence, we terminate training only when the batch-mean energy deviation satisfies $|E-E_0|<0.009$ and the maximum batch energy deviation satisfies $|E_{\max}-E_0|<0.03$. To reduce the effect of transient fluctuations, these two conditions must be satisfied in a cumulative total of 5 iterations.

\subsubsection{Characterization of the learned ground space}
After training, we sample $1500$ latent variables and decode them into circuit parameters to evaluate the learned ensemble. Among these generated states, $82.5\%$ satisfy both a relative energy error criterion $\Delta E = \frac{|E - (-2.5298)|}{|-2.5298|} < 0.02$ and a ground-space overlap score $p_G(\psi)=\sum_{i=1}^r |\langle g_i|\psi\rangle|>0.995$. This high success rate shows that the symmetry-encoded generative quantum circuit reliably prepares states within the AKLT ground space, even though the target states are represented through a symmetry-constrained parameterization rather than a conventional non-encoded ansatz.
\begin{figure}[htbp]
    \centering
    \includegraphics[width=0.6\linewidth]{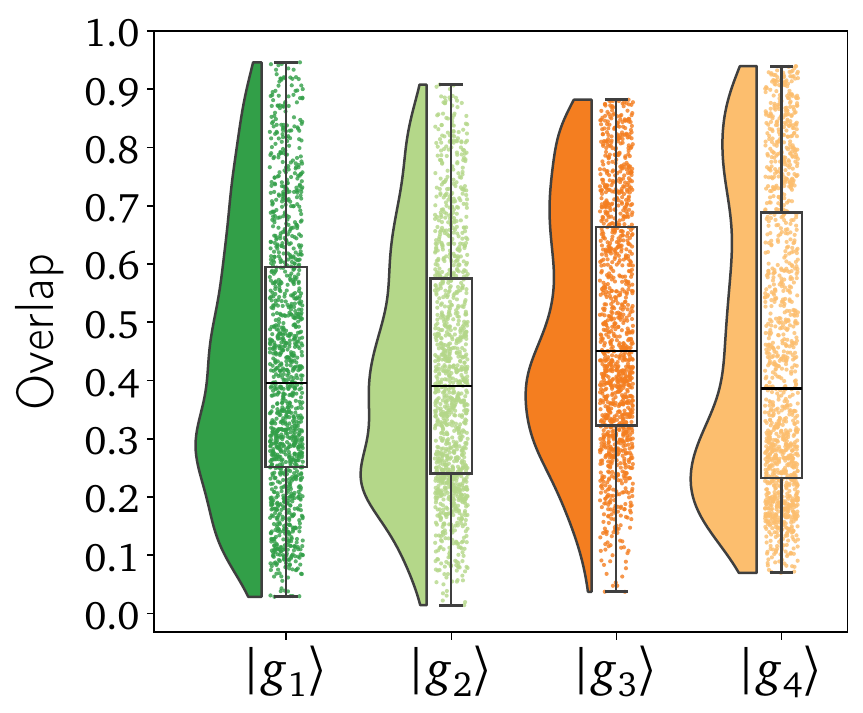}
  \caption{Overlap distributions of accepted states $\{|\psi_i\rangle\}_{i=1}^{1500}$ with the symmetry-encoded AKLT ground-state basis $\{|g_i\rangle\}_{i=1}^{4}$ at system size $N=10$. The half-violin plot represents the probability density, and the box-and-whisker plot summarizes the quartiles and spread. The broad distributions indicate that all basis directions are represented across the generated ensemble.}
  \label{fig:aklt-overlaps}
\end{figure}

To test whether these accepted states cover the full ground space, we examine the accepted-state matrix $C$ defined in Sec.~\ref{sec:Problem}. Using a tolerance $\varepsilon=0.05$, we recover the exact degeneracy, $\operatorname{rank}_{\varepsilon}(C)=\dim(\mathcal{H}_G)=4$. The associated singular values, $\{1,0.8737,0.8393,0.7607\}$, all remain well above the tolerance threshold, indicating that the generated states span the encoded AKLT ground space in a numerically stable way. This conclusion is consistent with the overlap distributions in Fig.~\ref{fig:aklt-overlaps}. For each basis state, the accepted ensemble contains samples with broadly distributed overlaps rather than being concentrated on only a few representative states. The learned ensemble therefore explores the different directions of the encoded AKLT ground space in a relatively balanced manner. 

The geometric agreement between the generated span and the exact ground space is further quantified by the principal angles $(\bar\alpha, \alpha_{\max})$ and chordal distance $d_{\mathrm{ch}}^2$ reported in Table~\ref{tab:metrics}. Both the mean and maximum principal angles are very small, and the corresponding chordal distance is close to zero. These metrics show that the space learned by the generative quantum circuit closely matches the exact encoded AKLT ground space $\mathcal{H}_G$.

A further notable feature of the learned ensemble is that it contains a subset of four generated states that together provide an approximate basis of the encoded AKLT ground space. Appendix~\ref{app:aklt} shows that these states reproduce the expected boundary magnetizations and edge-localized correlations associated with the unpaired edge spins. In this sense, the generative quantum circuit captures not only the correct ground-space span but also the physically meaningful basis structure of the encoded AKLT model.

\subsection{The spin-1 XXZ model}
The Hamiltonian of a spin-1 XXZ model under OBC is given by
\begin{equation}
    H_{\mathrm{XXZ}} = \sum_{i=1}^{N-1} \left( S_i^x S_{i+1}^x + S_i^y S_{i+1}^y + \Delta  S_i^z S_{i+1}^z \right),
\end{equation}
where $S_i^{\alpha}$ ($\alpha = x,y,z$) are spin-1 operators at site $i$, and $\Delta$ characterizes the anisotropy of the spin-exchange interaction in the model. At $\Delta = -1$, the system undergoes a first-order quantum phase transition~\cite{zhang2020quantum}. For $N'=4$ qutrits, the ground space exhibits a 9-fold degeneracy, with the ground states corresponding to the ferromagnetic multiplet.

\subsubsection{Energy-diversity objective and training dynamics}
As in the AKLT case, we adopt a symmetry-encoding scheme to represent the spin-1 XXZ chain on a variational qubit circuit (see Appendix~\ref{app:spin-ansatze}). In the present setting, the $N'=4$ spin-1 system is encoded into an $N=8$ qubit circuit, with exact ground-state energy $E_0=-3$. The classical generator uses a six-layer encoder and decoder with latent dimension $d_z=50$, and the variational circuit has depth $L=6$. Training is performed with the Adam optimizer, starting from a learning rate $\eta=6\times10^{-4}$ that is gradually decayed during optimization. The full set of hyperparameters is summarized in Table~\ref{tab:configuration}. To distinguish the degenerate ground states, we use the one-body feature vector $\bm{l}^{(1)}$ together with the nearest-neighbor two-body correlators $\bm{l}^{(2)}$, defined in Eq.~(\ref{eq:one/twocorre}).

\begin{figure}[htbp]
  \centering
    \includegraphics[width=0.65\linewidth]{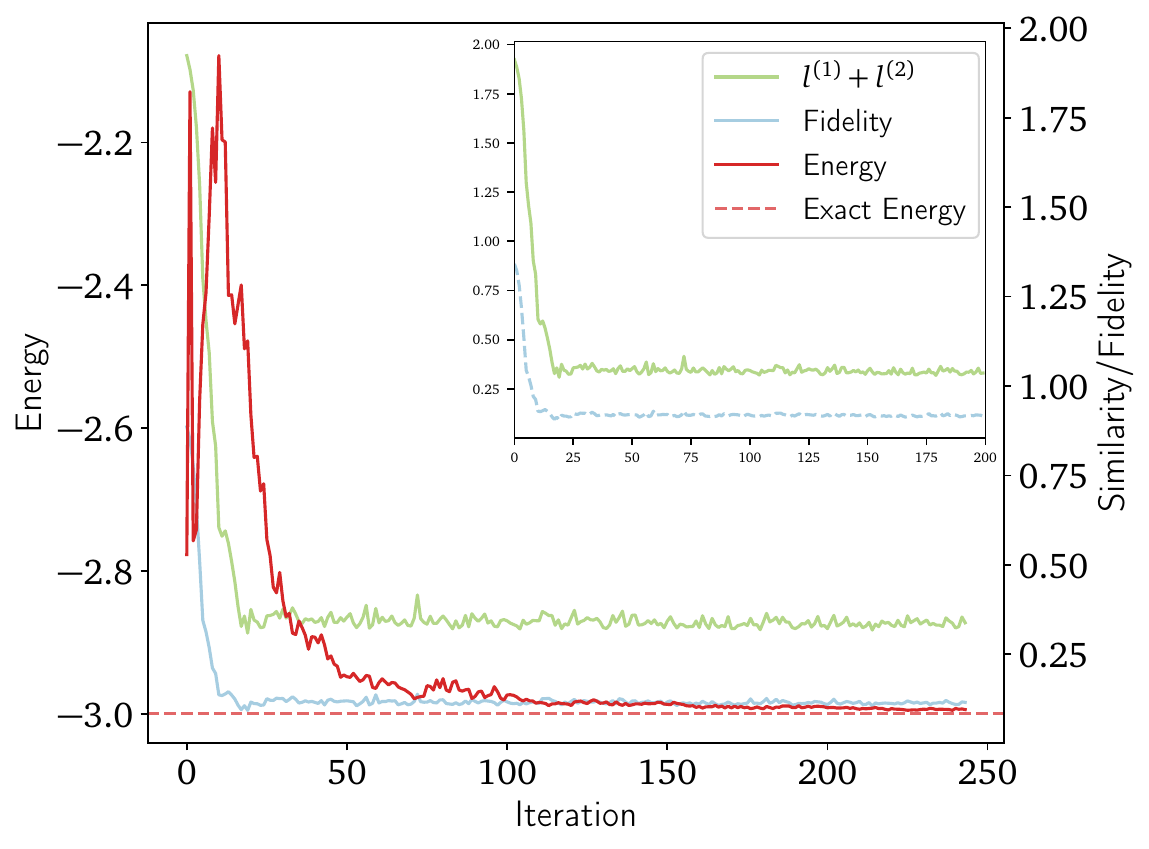}
  \caption{Training dynamics for the symmetry-encoded XXZ model at system size $N=8$. Shown are the batch-mean energy (red), batch-mean pairwise fidelity (blue), and the sum of the cosine similarities $l^{(1)} + l^{(2)}$ (green). The inset highlights the first 200 iterations. After an initial diversity-driven spreading stage with a rebound in the energy, the ensemble concentrates smoothly into the full degenerate ground space, with only weak late-stage fluctuation in similarity and fidelity.}
  \label{fig:xxz-dynamics}
\end{figure}

The training dynamics shown in Fig.~\ref{fig:xxz-dynamics} follow the same overall two-stage pattern observed in the MG and AKLT models, but it exhibits a stronger competition at early times, and a more stable convergence once the ensemble has entered the degenerate ground space. At the beginning of training, the diversity weight is set to $\lambda_2=1.5$, which drives the generated states to spread rapidly across Hilbert space. This is reflected in the sharp initial drop of the batch-mean pairwise fidelity. During this early stage, the batch-mean energy exhibits a noticeable nonmonotonic evolution, including a clear rebound, indicating that the rapid spreading of the ensemble initially competes strongly with energy minimization. As $\lambda_2$ is annealed to $0.4$, the energy term becomes dominant and drives the batch-mean energy smoothly toward the exact value $E_0=-3$. In contrast to the MG and AKLT cases, once the ensemble enters the ground space, the subsequent evolution shows only weak fluctuation in similarity and fidelity. This indicates that, after the initial transient stage, the generated states organize efficiently within the degenerate ground space, so that the later optimization involves a weaker competition between enforcing diversity and lowering the energy.

Finally, we terminate the optimization only when the batch-mean energy deviation satisfies $|E - E_0| < 0.01$ and the maximum batch energy deviation satisfies $|E_{\max} - E_0| < 0.02$. To reduce the influence of fluctuations, these two conditions must be satisfied in a cumulative total of 5 iterations.

\subsubsection{Characterization of the learned ground space}
We decode and execute $1500$ latent samples to evaluate the learned ensemble. Among these generated states, $92.7\%$ satisfy the acceptance criteria, namely a relative energy error $\Delta E = \frac{|E - (-3)|}{|-3|}<0.017$, together with a ground-space overlap score $p_G(\psi)>0.995$. This high acceptance rate indicates that the generative quantum circuit reliably produces ground states inside the degenerate ground space of the symmetry-encoded XXZ model.

\begin{figure}[htbp]
    \centering
    \includegraphics[width=0.6\linewidth]{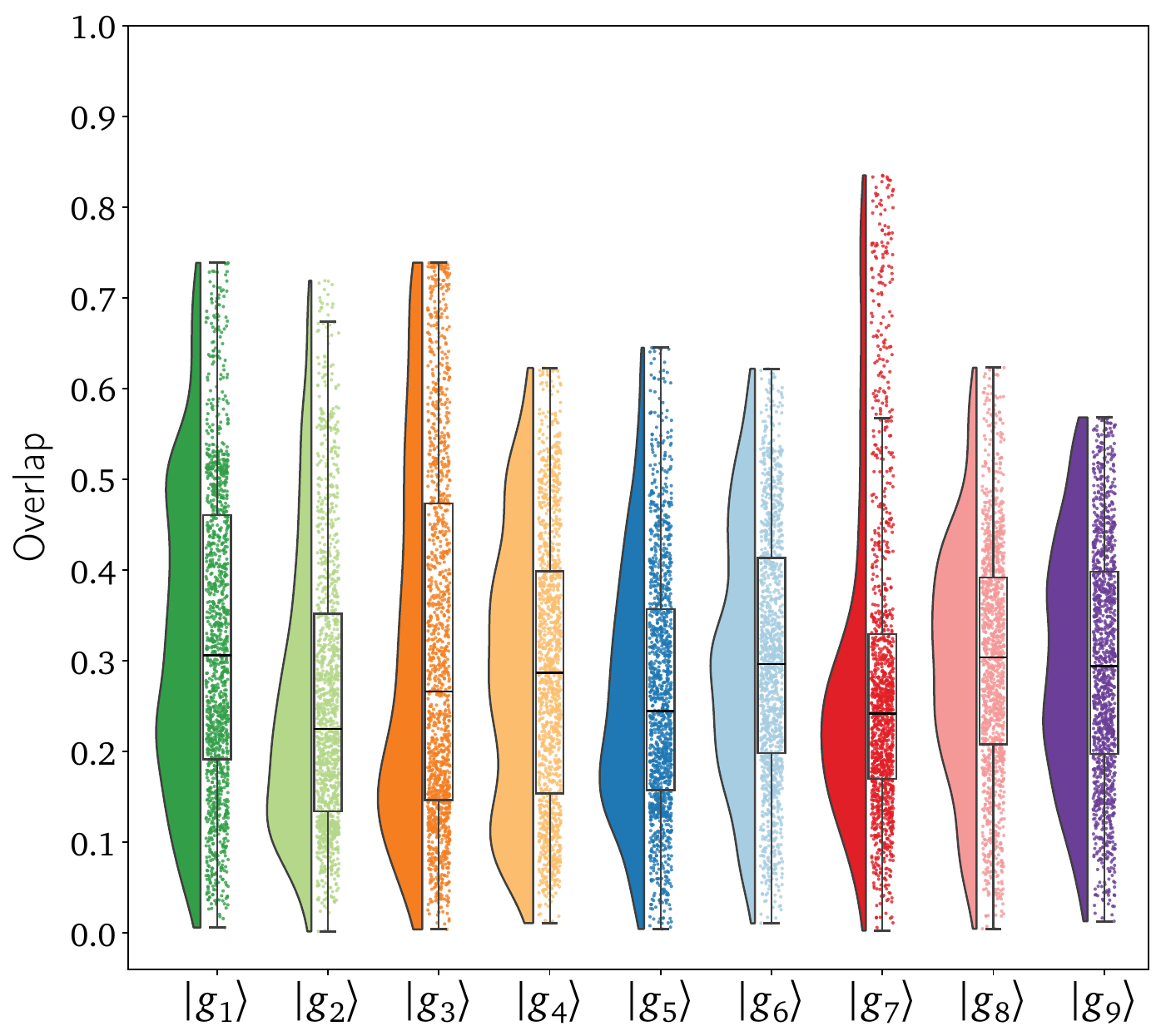}
  \caption{Overlap distributions of accepted states $\{|\psi_i\rangle\}_{i=1}^{1500}$ with the symmetry-encoded XXZ ground-state basis $\{|g_i\rangle\}_{i=1}^{9}$ at system size $N=8$.  The plots combine probability densities (the half-violin plot on the left) with quartile statistics (the box-and-whisker plot on the right box). The broad distributions spanning the range $[0,1]$ indicate a diverse generated ensemble.}
  \label{fig:xxz-overlaps}
\end{figure}

We analyze the coefficient matrix $C$ defined in Sec.~\ref{sec:Problem} to show that these accepted states recover the full ground space. The numerical rank is $\operatorname{rank}_{\varepsilon}(C)=9$ at a tolerance of $\varepsilon=0.03$, exactly matching the degeneracy of the XXZ model. The associated singular values, $\{1, 0.9099, 0.8442, 0.8201, 0.8084, 0.7799, 0.7586, 0.7239, \\ 0.7138\}$, remain well above the tolerance threshold, confirming that the accepted ensemble has populated all linearly independent ground-state directions in the target space. 

This space coverage is also reflected directly in the sample distribution. Figure~\ref{fig:xxz-overlaps} shows the overlap distribution over the exact symmetry-encoded basis. For each basis state, the overlaps are broadly distributed. The generated ensemble therefore explores the different directions of the ground space in a comparatively balanced way, instead of favoring only a small subset of representative states. Consistently, the principal angles $(\bar\alpha,\alpha_{\max})$ are very small and the corresponding chordal distance satisfies $d_{\mathrm{ch}}^2\approx 0$, as reported in Table~\ref{tab:metrics}. Taken together, these metrics indicate that the generated span is numerically indistinguishable from the exact ground space $\mathcal{H}_G$.

Having established the correct space coverage, we further examine whether the feature-space structure is also reproduced. To this end, we select $10$ representative generated states and compare their cosine-similarity matrices with those of the exact ground-state basis (see Appendix~\ref{app:xxz}). The resulting patterns are in close agreement, showing that the generated states faithfully reproduce the characteristic correlations of the exact XXZ ground states. This result further demonstrates that the local-observable-based diversity term remains effective even when the degeneracy is associated with a critical point, enabling the VGON to capture not only the correct ground-space span but also its nontrivial feature structure.

\section{Discussion}

In this work, we introduce a local-observable-guided generative quantum-circuit framework for learning degenerate ground spaces of quantum many-body Hamiltonians. By combining energy minimization with a diversity term constructed from local one-body and two-body observables, the method generates an ensemble of states whose span recovers the full degenerate ground space, while avoiding the cost of full tomography.

We benchmark the approach on three representative quantum spin models with distinct origins of degeneracy: the Majumdar-Ghosh model, the Affleck-Kennedy-Lieb-Tasaki model, and the spin-1 XXZ chain at its first-order transition point. Across all cases, the generated ensembles achieve high acceptance rates under strict energy and overlap criteria, reproduce the exact degeneracy through the tolerance rank of the accepted-state matrix, and show close agreement with the exact ground space as quantified by principal angles and chordal distance. In addition, feature-space analyses show that the generated states capture not only the correct ground-space span but also the internal structure of the target ground space. We further show, through a finite-shot benchmark on the MG model, that the framework remains effective under shot-based estimation and can still recover the degenerate ground space with a reduced measurement budget.

These results establish local-observable-guided generative quantum circuits as a practical framework for learning degenerate ground spaces. An important direction for future work is to extend the method to more general topological or symmetry-protected phases, where the design of informative local features may be less direct. More broadly, the present results suggest that combining generative modeling with physically motivated observables can provide a useful route toward larger-scale quantum many-body state discovery beyond single-state variational optimization.

\section*{Acknowledgments}
This work was supported by the NSFC (No.~12574536) and the Sichuan Science and Technology Program (2024YFHZ0371).
\bibliography{refs}
\appendix

\section{Ansatz Design for Degenerate Ground Spaces}
\label{app:ansatz}

In this appendix, we present the parameterized quantum-circuit ans\"atze used to represent degenerate ground spaces in the models considered in the main text. Since the spin-$\tfrac{1}{2}$ and spin-1 systems differ substantially in their local Hilbert-space structure and symmetry constraints, we use two distinct circuit constructions. The first is tailored to the spin-$\tfrac{1}{2}$ MG model, and the second is built for the spin-1 AKLT and XXZ models using a symmetry-encoding scheme. These ans\"atze are designed to balance expressiveness and structural compatibility, so that the corresponding circuits can represent the relevant degenerate ground-state manifolds efficiently.

\subsection{Circuit ansatz for spin-$\frac{1}{2}$ model}
\label{app:mg-ansatze}
\begin{figure}[htbp]
  \centering
  \begin{subfigure}[t]{0.47\textwidth}
    \centering
    \includegraphics[width=\linewidth]{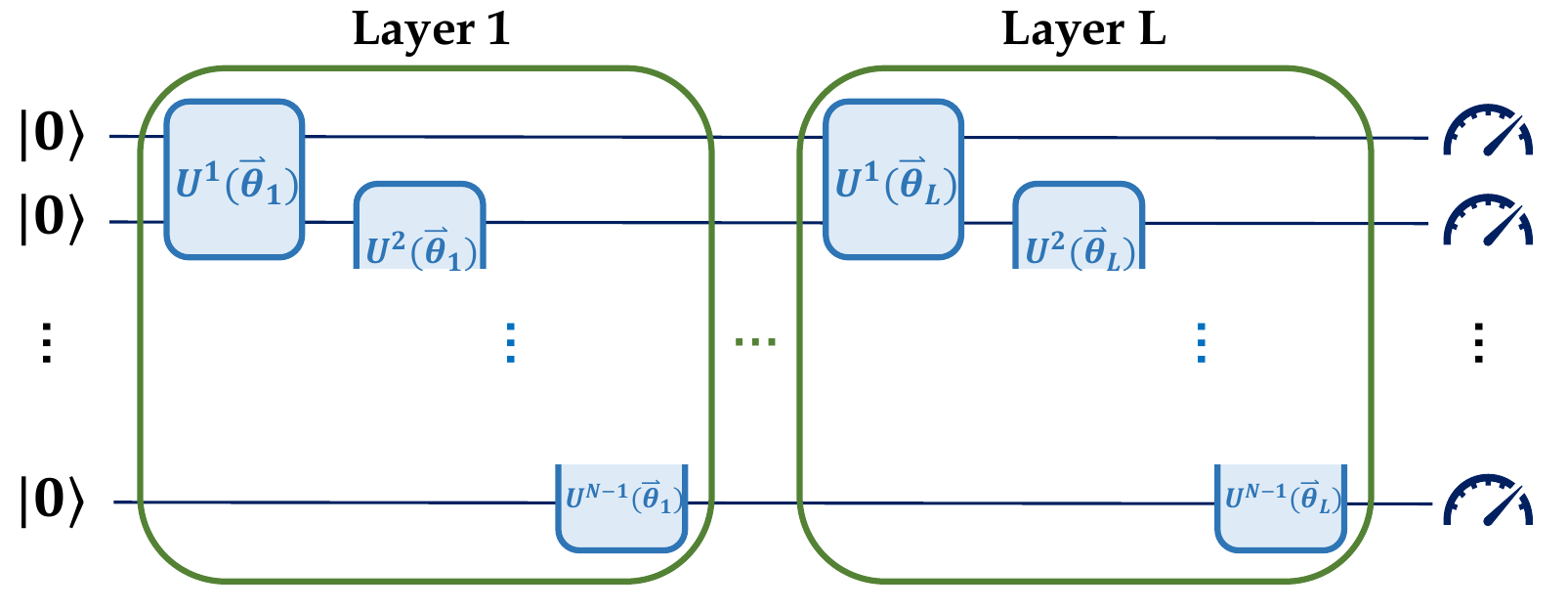}
    \caption{Each green box represents a layer of the circuit consisting of $n-1$ universal 2-qubit gate blocks} 
  \end{subfigure}
  \hspace{0.1mm}
  \begin{subfigure}[t]{0.47\textwidth}
    \centering
    \includegraphics[width=\linewidth]{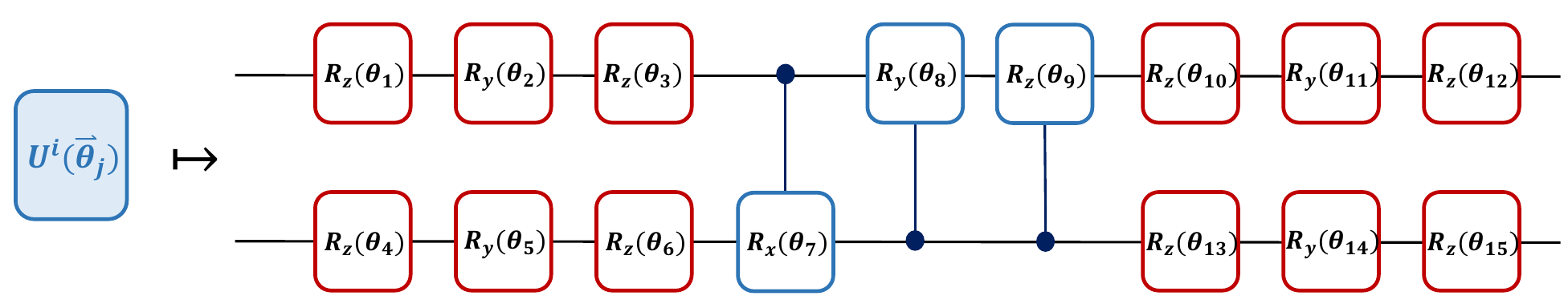}
    \caption{Each universal 2-qubit gate block is decomposed into 12 rotation gates and 3 controlled rotation gates} 
  \end{subfigure}

  \caption{Structure of the circuit ansatz for spin-$\frac{1}{2}$ model consisting of $L$ layers.}
  \label{fig:mg-ans\"atze}
\end{figure}

To represent the degenerate ground space of the Majumdar-Ghosh (MG) model, we employ a variational ansatz built from sequential layers of nearest-neighbor two-qubit gates. As shown in Fig.~\ref{fig:mg-ans\"atze}, each circuit layer consists of a chain of universal two-qubit gate blocks, and each block is composed of 6 layers of single-qubit rotations together with 3 controlled one-parameter rotation gates. This construction is adapted from the framework of Ref.~\cite{zhangVariationalOptimizationQuantum2025}, but replaces fixed CNOT gates by tunable controlled rotations in order to improve the flexibility of the circuit.

The motivation for this modification is that fixed entangling gates are too restrictive for representing the correlations required by the degenerate ground space. In practice, such rigidity makes the generative optimization less stable, since latent samples may be mapped to states that are widely dispersed in Hilbert space rather than concentrated near the target ground space. By replacing CNOT gates with controlled one-parameter rotations, the circuit acquires additional continuous degrees of freedom, which enables smoother changes in the generated states during training. This improves the stability of the optimization, while maintaining sufficient expressiveness to capture the relevant degenerate ground-state structure.

\subsection{ Circuit ansatz for spin-1 models}
\label{app:spin-ansatze}

For the spin-1 models, we employ a symmetry-based encoding scheme in which the Hilbert space of $N'$ spin-1 degrees of freedom is embedded into the Hilbert space of $N=2N'$ qubits. Each spin-1 site is represented in the symmetric subspace of a qubit pair, so that the encoded qubit circuit preserves the local structure and symmetries of the original spin-1 system~\cite{yangCostLocallyApproximating2025}.

\begin{figure}[htbp]
  \centering
    \includegraphics[width=\linewidth]{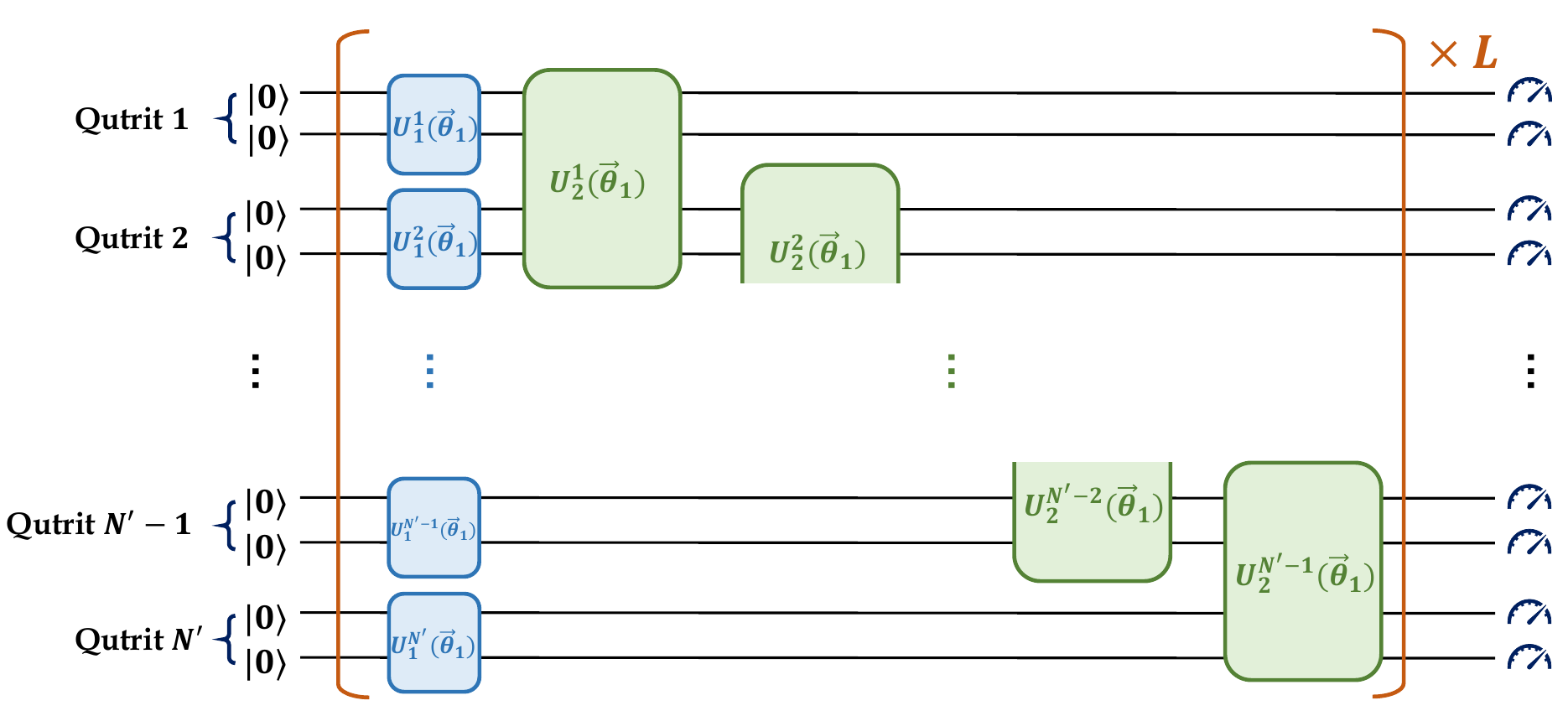}
  \caption{Circuit ansatz for the spin-$1$ model, consisting of $L$ layers. Each qutrit is encoded using two qubits. The orange bracket denotes one circuit layer, composed of $N'$ single-qutrit unitary gates (blue blocks) and $N'-1$ controlled-diagonal synthesis gates (green blocks). }
  \label{fig:spin1-ans\"atze}
\end{figure}
Within this encoding, the ansatz is built entirely from symmetry-preserving gates, so that the variational dynamics remain confined to the encoded subspace and leakage into unphysical states is avoided. As illustrated in Fig.~\ref{fig:spin1-ans\"atze}, each layer consists of local single-qutrit unitaries acting on every encoded site together with controlled-diagonal synthesis gates on nearest-neighbor pairs. Rather than adopting arbitrary two-qutrit unitaries with a much larger parameter count~\cite{yangCostLocallyApproximating2025}, we use this controlled-diagonal construction to obtain a substantially more economical encoded circuit~\cite{PhysRevA.87.012325}, while preserving the expressiveness needed to capture the relevant spin-1 ground-state structure. This compact architecture is sufficiently general to describe both the AKLT model and the spin-1 XXZ model.

To simulate a spin-1 Hamiltonian $\tilde{H}$ on a qubit processor, we encode the effective spin-1 operators $\mathbf{S}'$ into qubit operators acting on each qubit pair according to
\begin{equation}
    \begin{split}
        \mathbf{S}'_x &= \frac{S_x \otimes \mathcal{I} + \mathcal{I} \otimes S_x}{2} \\
        \mathbf{S}'_y &= \frac{S_y \otimes \mathcal{I} + \mathcal{I} \otimes S_y}{2} \\
        \mathbf{S}'_z &= \frac{S_z \otimes \mathcal{I} + \mathcal{I} \otimes S_z}{2} \\
    \end{split}
\end{equation}
This mapping yields an encoded qubit Hamiltonian $H$ that preserves the symmetries of the original spin-1 problem. The combination of the encoded Hamiltonian and the symmetry-preserving ansatz ensures that the variational circuit can effectively simulate the physical spin-1 space, while retaining sufficient expressive power to represent the relevant degenerate ground states.

\section{Cosine-Similarity Analysis in Local-Observable Space}
\label{app:cosine-local}

In this appendix, we further examine the structure of the generated ensemble in the local-observable feature space. The purpose is not only to confirm that the generated states span the correct ground space, but also to show that they reproduce the geometric organization of the exact ground-state basis in terms of physically meaningful local features. To this end, we select 10 representative states from the generated ensemble and project them onto the exact ground-state basis $\{|g_j\rangle\}_{j=1}^{k}$. Their overlap amplitudes, $w_{m,j} = \langle g_j | \psi_m \rangle$, provide a direct characterization of how these representative generated states are distributed among the exact ground-state directions.

We then compare the cosine-similarity matrices constructed from local feature vectors for the exact ground states and for the selected generated states. Specifically, the matrix elements are defined as
\begin{equation}
\bm{M}^{(k)}_{ij} = \frac{\langle \bm{l}^{(k)}(\bm{\theta}_i), \bm{l}^{(k)}(\bm{\theta}_j) \rangle}
{\| \bm{l}^{(k)}(\bm{\theta}_i) \| \| \bm{l}^{(k)}(\bm{\theta}_j) \|},
\end{equation}
where $\bm{l}^{(k)}(\bm{\theta})$ represents the local feature vector of type $k$, for example the vectors defined in Eq.~(\ref{eq:one/twocorre}) or Eq.~(\ref{eq:aklt-two-edge}) in the main text. These cosine-similarity matrices quantify the relative overlap of states using only local observables. By comparing the corresponding heatmaps, we show that the representative generated states reproduce a similar feature-space organization to the exact ground-state basis, indicating that the model captures its physically meaningful internal structure.

\subsection{MG model}
\label{app:mg}

For the Majumdar-Ghosh (MG) chain with open boundary conditions, which is defined as $H_{\mathrm{MG}}=\sum_{i=1}^{N-2} (\mathbf  S_i \cdot \mathbf S_{i+1} + \mathbf S_{i+1} \cdot \mathbf S_{i+2} + \mathbf S_i \cdot \mathbf S_{i+2} )$, ($\mathbf{S}_i$ are spin-$\tfrac{1}{2}$ operators), we consider system sizes $N=9$ and $N=10$, whose ground-space degeneracies are $k=4$ and $k=5$, respectively. 

\begin{figure}[htbp]
  \centering
  \begin{subfigure}[t]{0.32\textwidth}
    \centering
    \includegraphics[width=\linewidth]{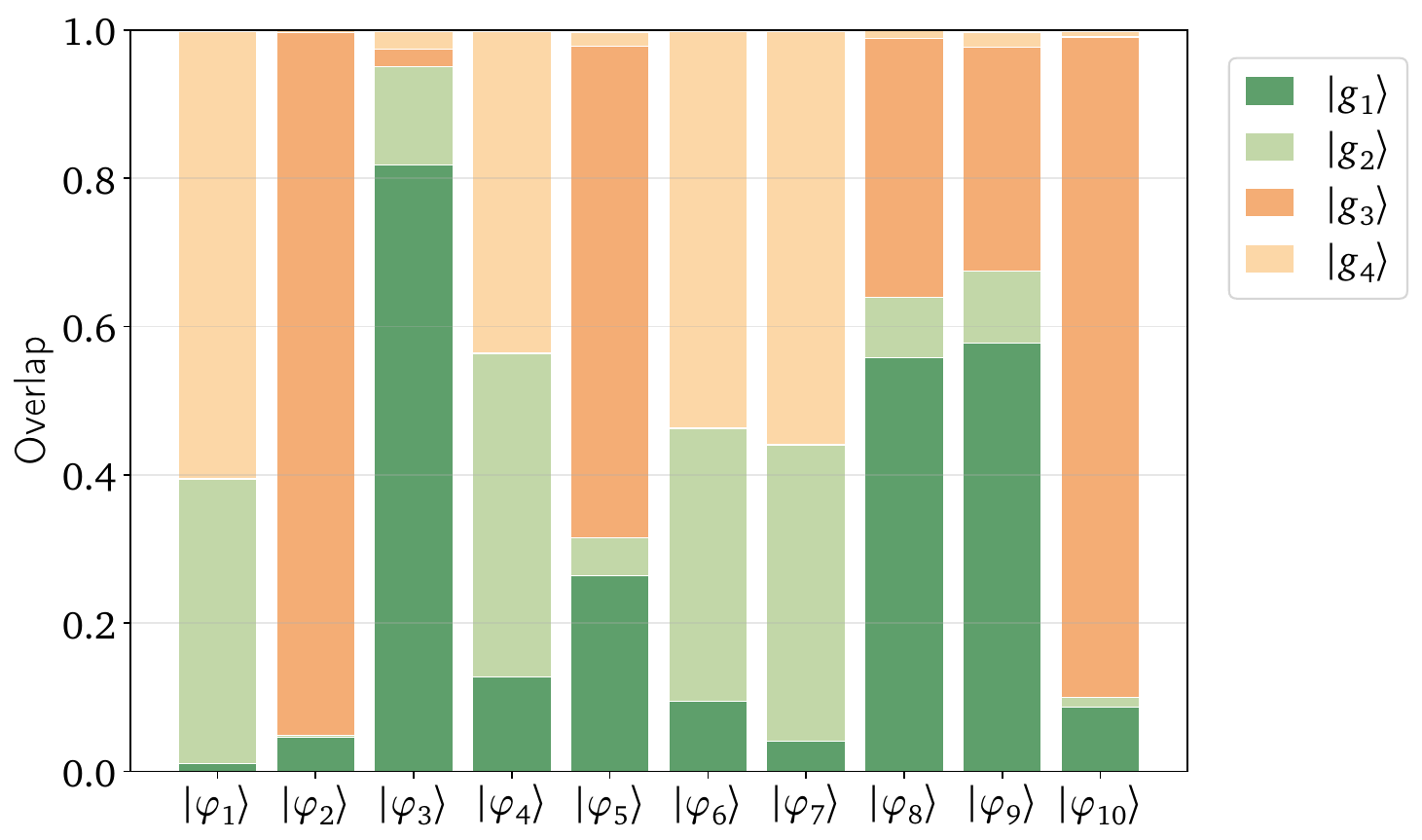}
    \caption{system size $N=9$ with ground-state degeneracy $k=4$}
  \end{subfigure}
  \hspace{0.1mm}
  \begin{subfigure}[t]{0.32\textwidth}
    \centering
    \includegraphics[width=\linewidth]{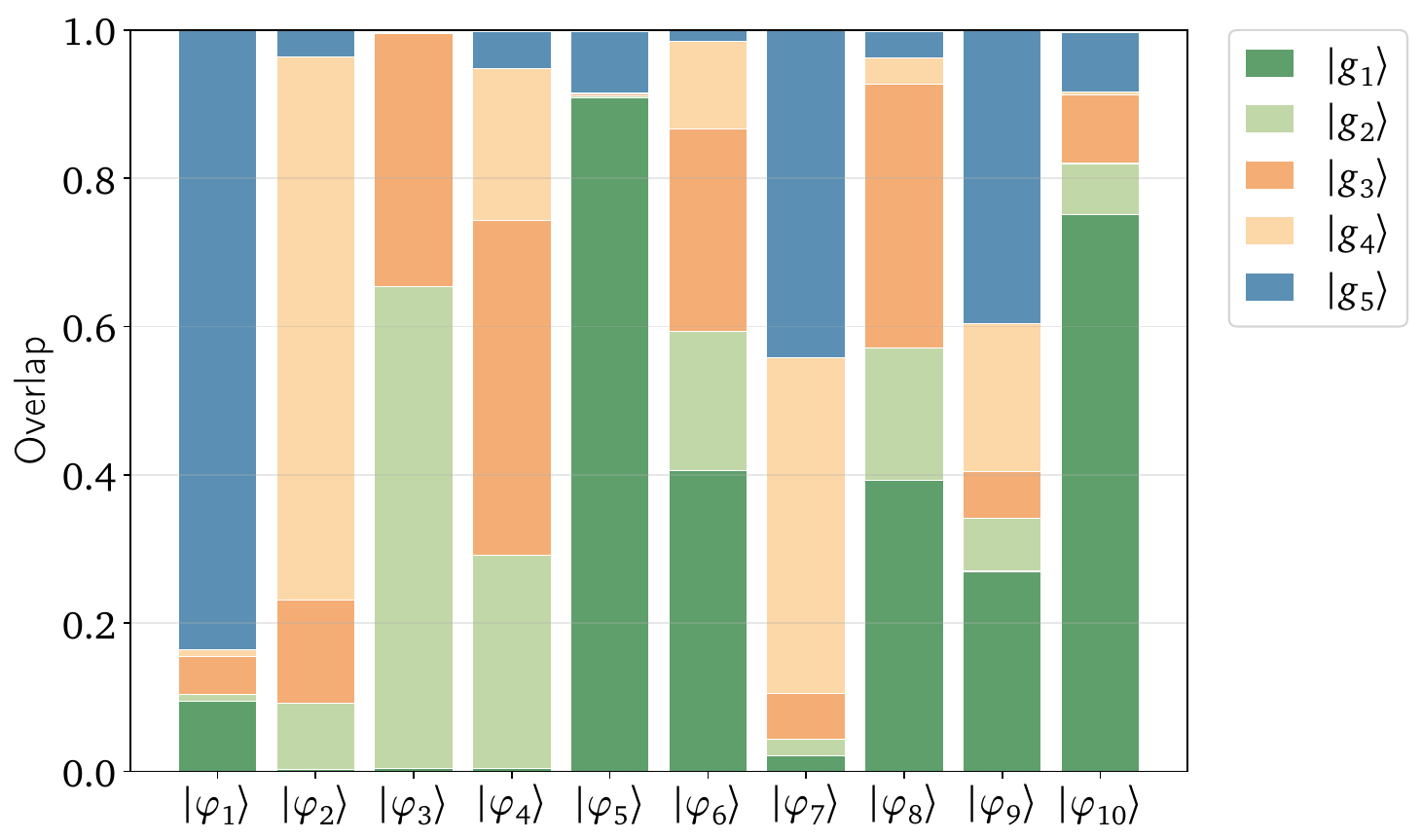}
    \caption{system size $N=10$ with ground-state degeneracy $k=5$}
  \end{subfigure}
  \caption{Overlaps of 10 representative generated states $\{|\psi_i\rangle\}_{i=1}^{10}$ with the exact MG ground-state basis $\{|g_i\rangle\}_{i=1}^{k}$. Each column shows the decomposition of a generated state in the exact basis. The first $k$ generated states form an approximately orthogonal subset.}
  \label{fig:app-mg-generate_states}
\end{figure}
Fig.~\ref{fig:app-mg-generate_states} shows the overlaps of 10 representative generated states with the exact MG ground-state basis. Each column corresponds to a generated state, and the color blocks indicate its decomposition over the exact basis vectors. The overlap patterns show that different generated states correspond to distinct linear combinations of the basis states, rather than repeatedly reproducing the same direction. In particular, among these representative states, one can identify approximately orthogonal subsets, i.e., $\{|\psi_i\rangle\}_{i=1}^{4}$ for $N=9$ and $\{|\psi_i\rangle\}_{i=1}^{5}$ for $N=10$. The mutual overlaps within these subsets are below $0.05$ for $N=9$ and below $0.1$ for $N=10$. This provides an explicit sample-level illustration that the generated ensemble contains states spanning the full MG ground space.

\begin{figure*}[htbp]
  \centering
  \begin{subfigure}[t]{0.23\textwidth}
    \centering
    \includegraphics[width=\linewidth]{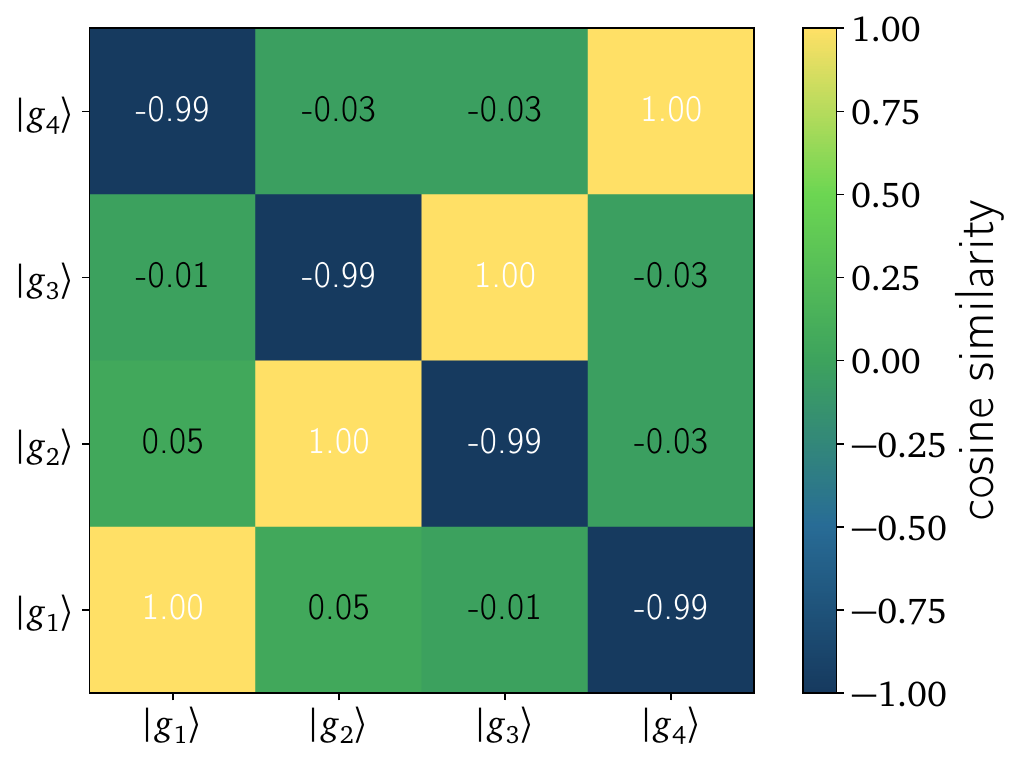}
    \caption{True states ($N=9, \bm{l}^{(1)}$)}
    \label{fig:mg-true-9-l1}
  \end{subfigure}
  \hfill
  \begin{subfigure}[t]{0.23\textwidth}
    \centering
    \includegraphics[width=\linewidth]{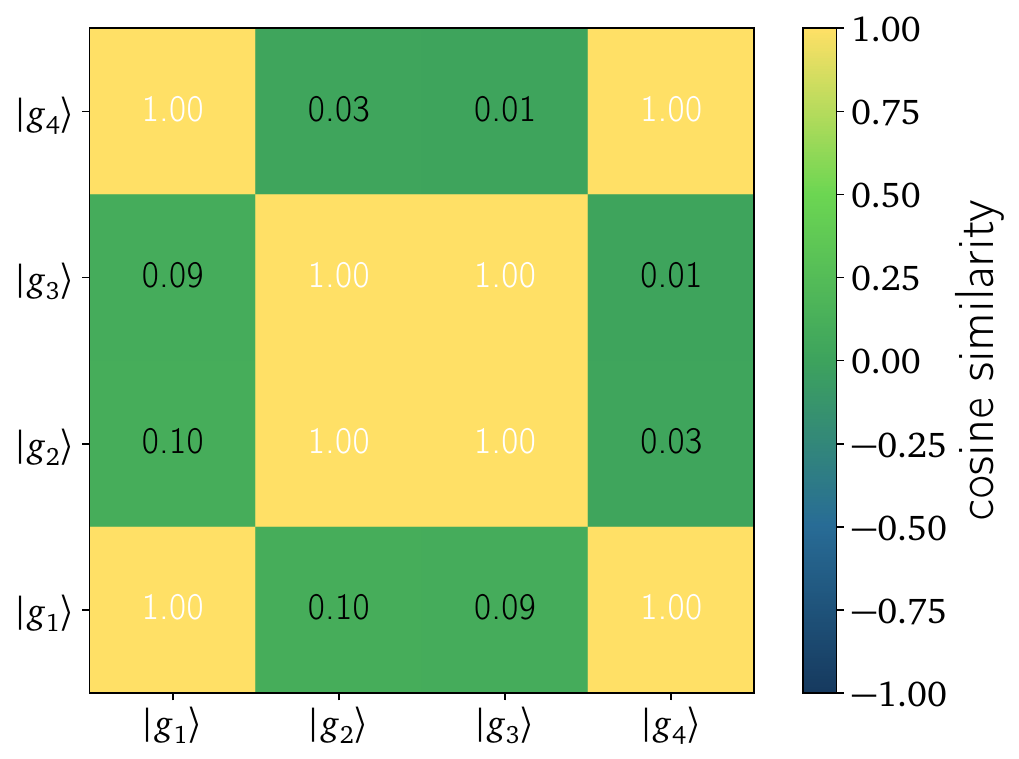}
    \caption{True states ($N=9, \bm{l}^{(2)}$)}
    \label{fig:mg-true-9-l2}
  \end{subfigure}
  \hfill
  \begin{subfigure}[t]{0.23\textwidth}
    \centering
    \includegraphics[width=\linewidth]{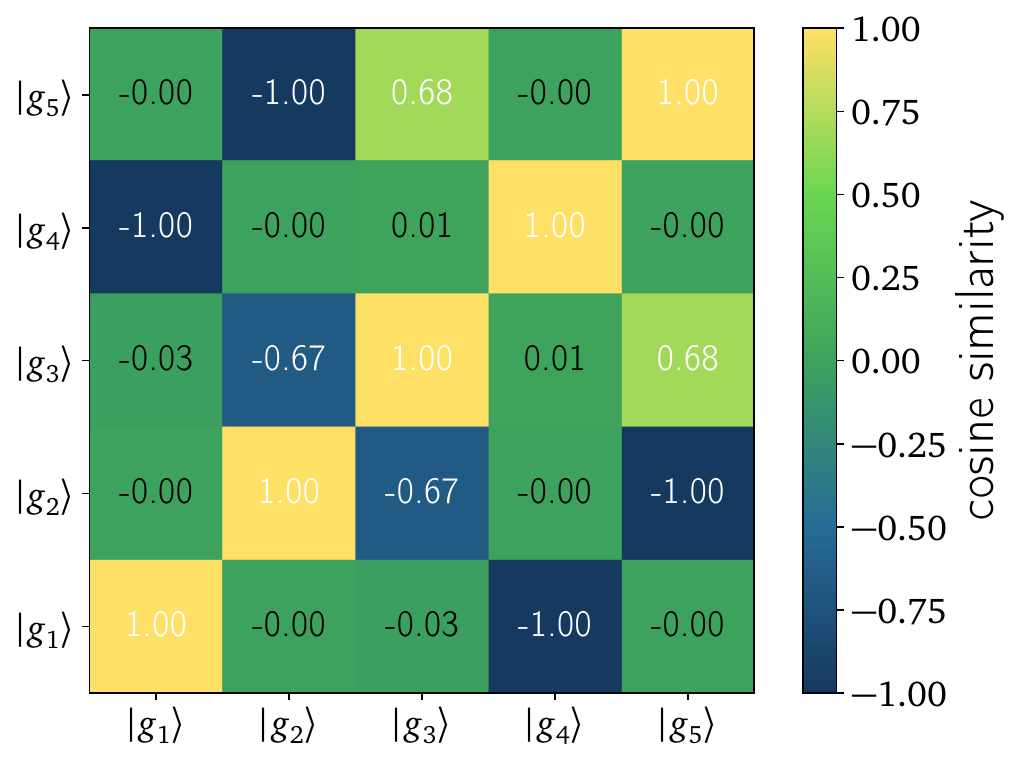}
    \caption{True states ($N=10, \bm{l}^{(1)}$)}
    \label{fig:mg-true-10-l1}
  \end{subfigure}
  \hfill
  \begin{subfigure}[t]{0.23\textwidth}
    \centering
    \includegraphics[width=\linewidth]{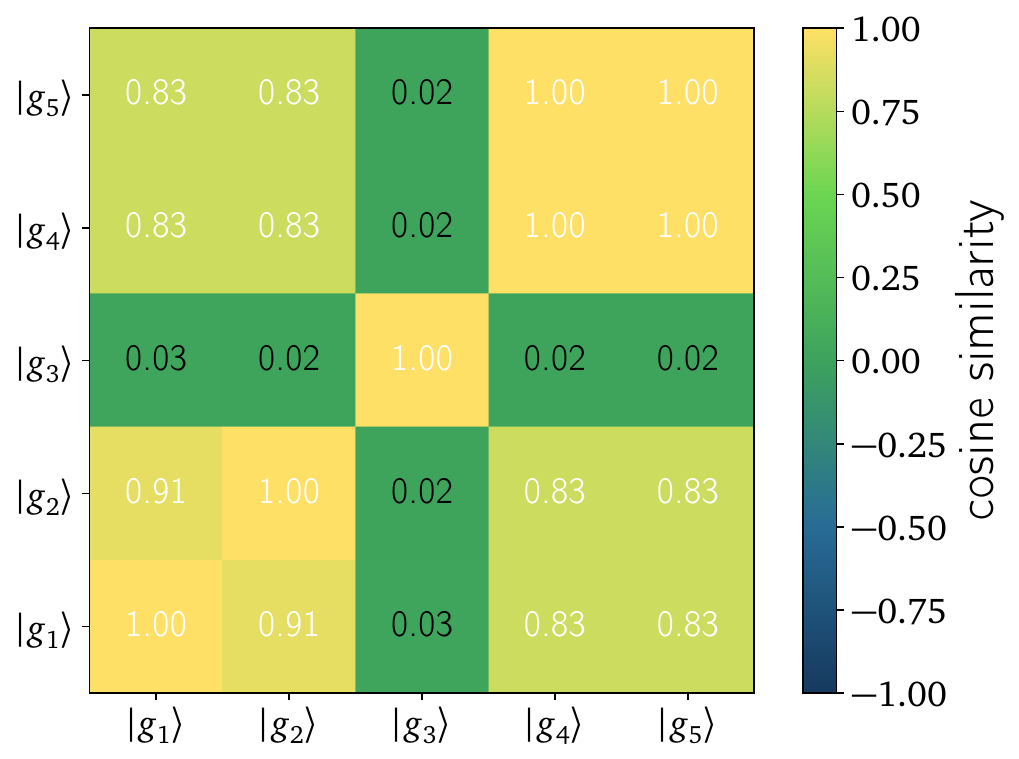}
    \caption{True states ($N=10, \bm{l}^{(2)}$)}
    \label{fig:mg-true-10-l2}
  \end{subfigure}

  \vspace{0.1mm}

  \begin{subfigure}[t]{0.23\textwidth}
    \centering
    \includegraphics[width=\linewidth]{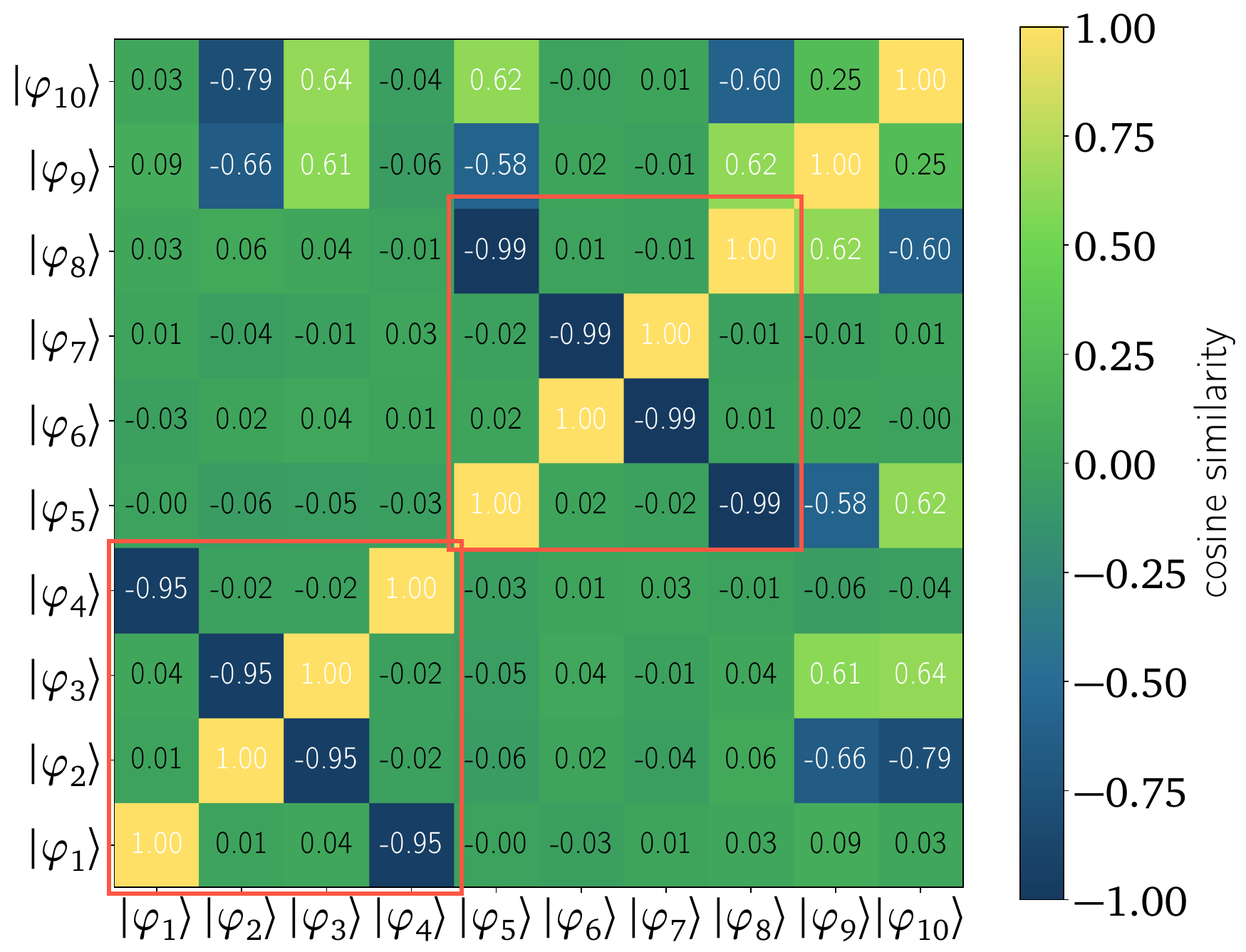}
    \caption{Generated states ($N=9, \bm{l}^{(1)}$)}
    \label{fig:mg-gen-9-l1}
  \end{subfigure}
  \hfill
  \begin{subfigure}[t]{0.23\textwidth}
    \centering
    \includegraphics[width=\linewidth]{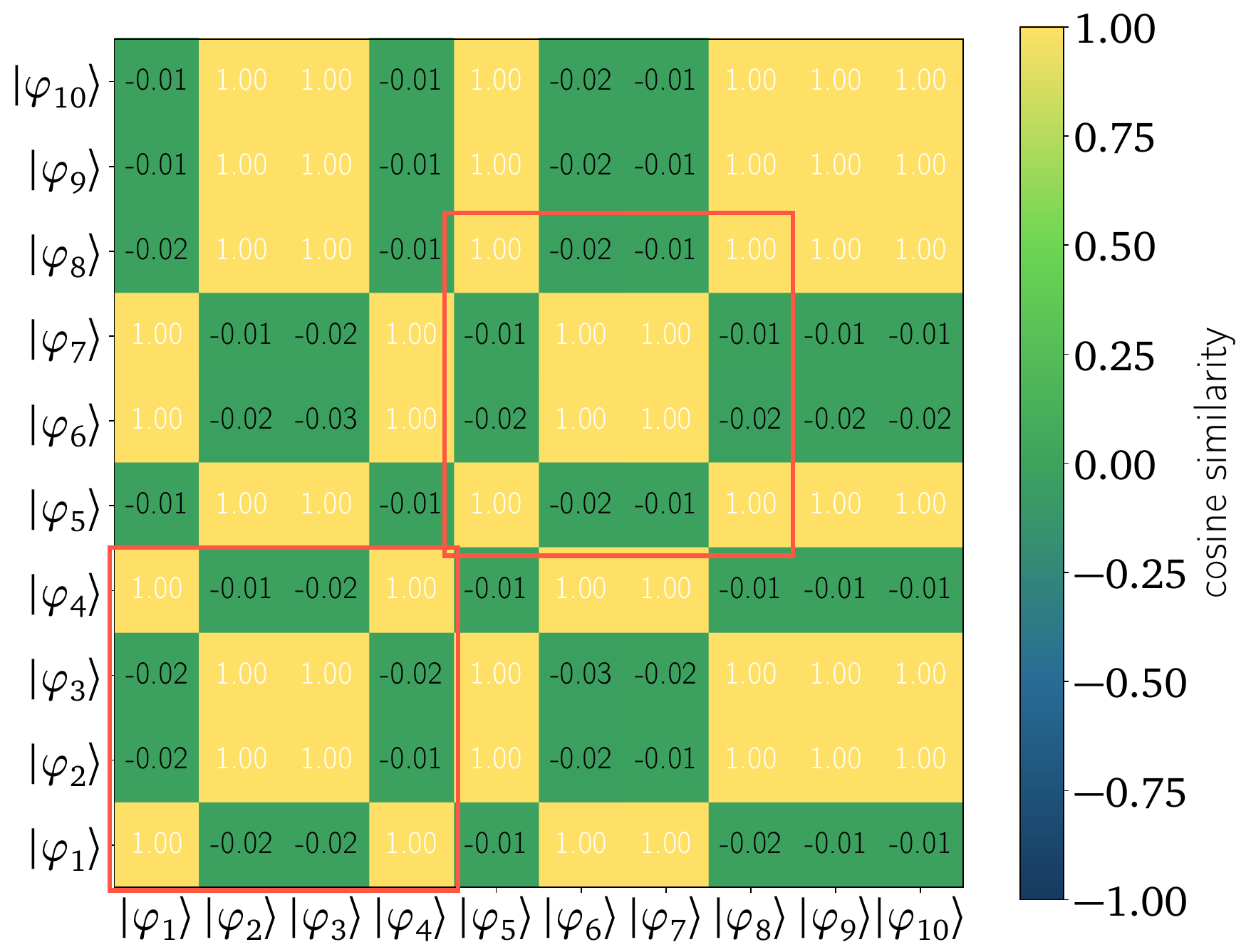}
    \caption{Generated states ($N=9, \bm{l}^{(2)}$)}
    \label{fig:mg-gen-9-l2}
  \end{subfigure}
  \hfill
  \begin{subfigure}[t]{0.23\textwidth}
    \centering
    \includegraphics[width=\linewidth]{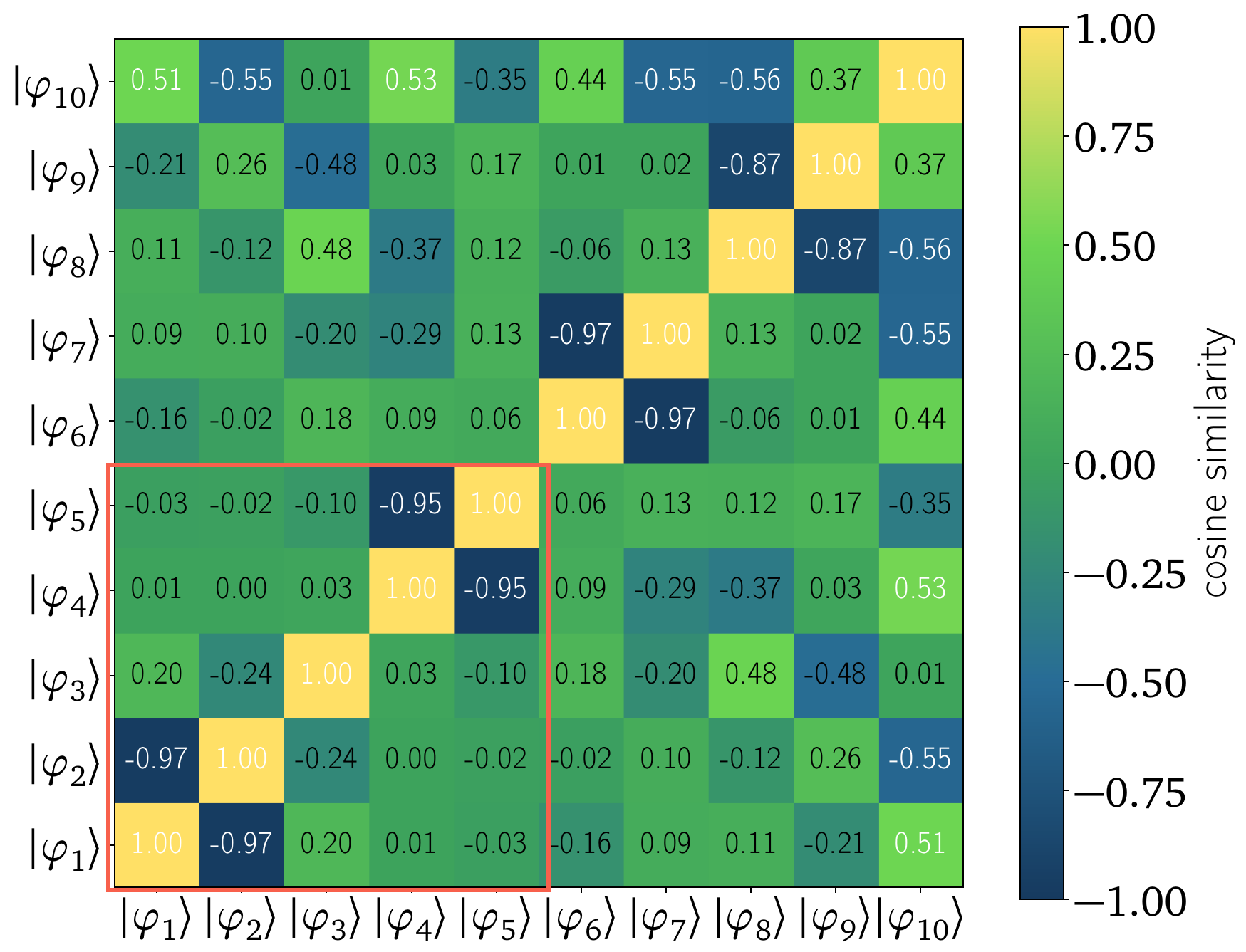}
    \caption{Generated states ($N=10, \bm{l}^{(1)}$)}
    \label{fig:mg-gen-10-l1}
  \end{subfigure}
  \hfill
  \begin{subfigure}[t]{0.23\textwidth}
    \centering
    \includegraphics[width=\linewidth]{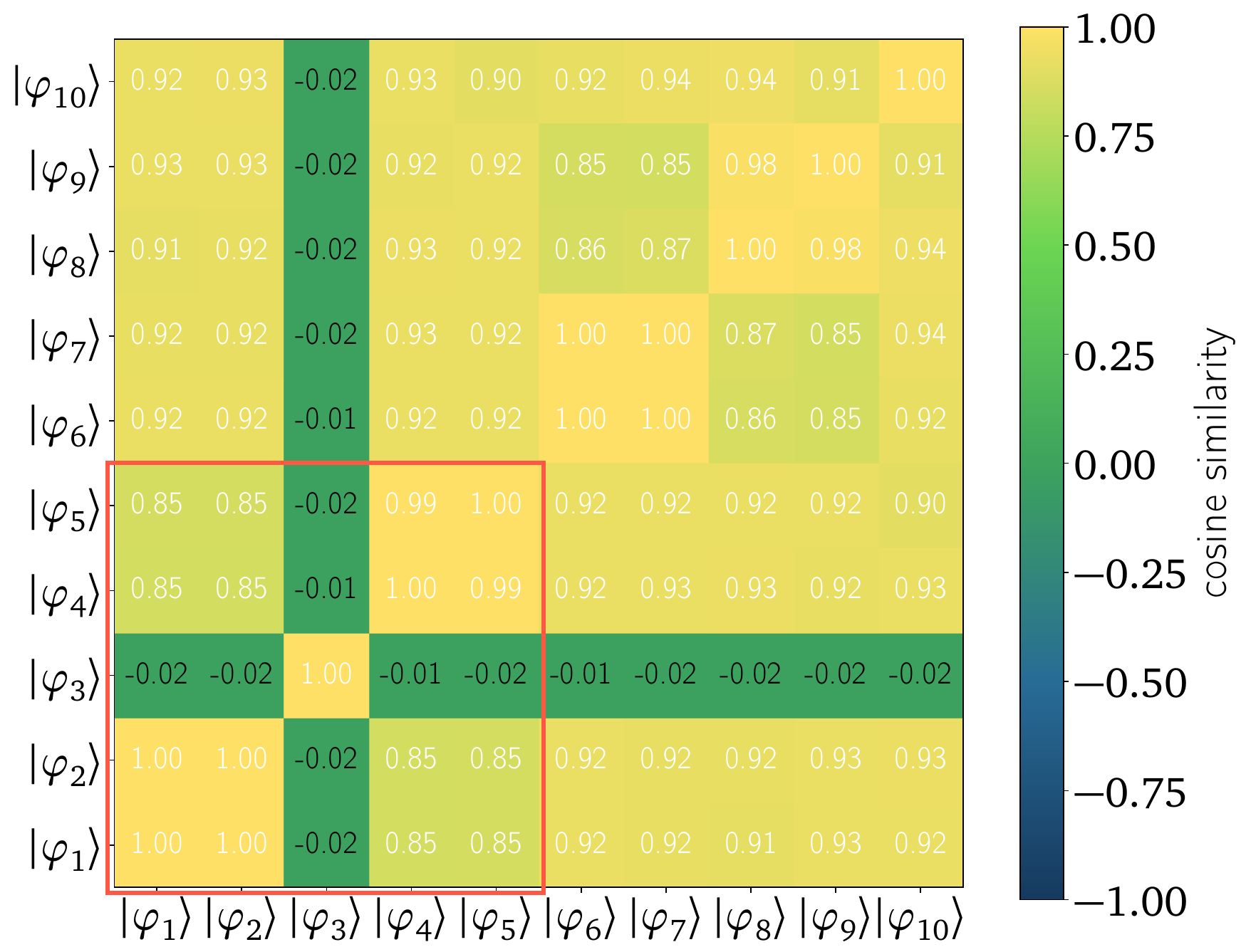}
    \caption{Generated states ($N=10, \bm{l}^{(2)}$)}
    \label{fig:mg-gen-10-l2}
  \end{subfigure}

  \caption{Cosine-similarity matrices for the MG model. The upper row (a)-(d) shows the exact ground-state basis $\{|g_i\rangle\}_{i=1}^{k}$, and the lower row (e)-(h) shows 10 representative generated states $\{|\psi_i\rangle\}_{i=1}^{10}$. Panels (a), (b), (e), (f) correspond to $N=9$, while panels (c), (d), (g), (h) correspond to $N=10$. For each system size, the matrices are computed from one-body expectation values $\bm{l}^{(1)}$ and nearest-neighbor two-body correlators $\bm{l}^{(2)}$. The red boxes highlight approximately orthogonal subsets of generated states whose similarity patterns match those of bases in the exact ground space.}
  \label{fig:mg-cosine-comparison}
\end{figure*}

To further examine the structure of the generated ensemble, Fig.~\ref{fig:mg-cosine-comparison} compares the cosine-similarity matrices of the exact ground states and the representative generated states. The upper row shows the exact basis, while the lower row shows the generated states. In each case, the matrices are computed separately from the one-body features $\bm{l}^{(1)}$ and the nearest-neighbor two-body correlators $\bm{l}^{(2)}$. The close correspondence between the upper and lower rows shows that the generated states reproduce the same feature-space organization as the exact MG ground states. The red boxes highlight approximately orthogonal subsets of generated states whose similarity structure matches that of true orthogonal bases.

For $N=9$, the exact cosine-similarity matrices exhibit a clear block structure controlled by the sensitivities of the two feature sets. In the $\bm{l}^{(2)}$ representation, which probes the dimer pattern, states sharing the same dimer background are indistinguishable. Concretely, the pairs $|g_1\rangle$ and $|g_4\rangle$, as well as $|g_2\rangle$ and $|g_3\rangle$, have cosine similarity equal to $1$ in $\bm{l}^{(2)}$ (Fig.~\ref{fig:mg-true-9-l2}). This degeneracy is lifted by the one-body features $\bm{l}^{(1)}$, where the same pairs become strongly anticorrelated ($\bm{l}^{(1)} \approx -1$, Fig.~\ref{fig:mg-true-9-l1}). States associated with different dimer coverings, by contrast, remain nearly orthogonal in both feature spaces.

For $N=10$, the structure is different but the same separation mechanism persists. In the $\bm{l}^{(2)}$ representation, $|g_3\rangle$ has a distinct dimer configuration and is nearly orthogonal to the remaining states ($\bm{l}^{(2)} \approx 0$, Fig.~\ref{fig:mg-true-10-l2}), whereas $|g_1\rangle,|g_2\rangle,|g_4\rangle,$ and $|g_5\rangle$ share similar dimer features and therefore display large cosine similarities. These residual degeneracies are again resolved by the one-body features $\bm{l}^{(1)}$, where the pairs $(|g_1\rangle,|g_4\rangle)$ and $(|g_2\rangle,|g_5\rangle)$ become anticorrelated ($\bm{l}^{(1)} \approx -1$), while the remaining cross terms stay close to orthogonal [Fig.~\ref{fig:mg-true-10-l1}].

The generated states reproduce these same structural patterns. For $N=9$, the cosine-similarity matrices in Figs.~\ref{fig:mg-gen-9-l1} and \ref{fig:mg-gen-9-l2} recover the same block organization as the exact basis. States sharing a dimer background remain nearly identical in $\bm{l}^{(2)}$, while $\bm{l}^{(1)}$ separates them through opposite local magnetization patterns ($\bm{l}^{(1)} \approx -1$). For $N=10$, the generated ensemble likewise reproduces the coexistence of high-similarity states in $\bm{l}^{(2)}$ and their separation in $\bm{l}^{(1)}$ (Figs.~\ref{fig:mg-gen-10-l1} and ~\ref{fig:mg-gen-10-l2}). These comparisons show that the generative quantum circuit does not merely reproduce the correct span of the MG ground space, it also captures the physically meaningful organization by identifying dimer backgrounds through $\bm{l}^{(2)}$ and resolving the remaining degeneracies through $\bm{l}^{(1)}$.

\subsection{AKLT model}
\label{app:aklt}

For the symmetry-encoded AKLT chain at system size $N=10$, which is defined as $H_{\mathrm{AKLT}} =\sum_{i=1}^{N-1} \left(\mathbf S_i \cdot \mathbf S_{i+1}
+\tfrac{1}{3} (\mathbf S_i \cdot \mathbf S_{i+1})^2\right)$ ($\mathbf S_i$ are spin-1 operators on site $i$), the fourfold ground-state degeneracy originates from the fractionalized edge spin-$\tfrac12$ degrees of freedom. Accordingly, the most informative observables are localized near the boundaries. In this appendix, we further examine how representative generated states are distributed with respect to the exact ground-state basis, and whether the generated ensemble reproduces the feature-space organization associated with the AKLT edge structure.

\begin{figure}[htbp]
    \centering
    \includegraphics[width=0.7\linewidth]{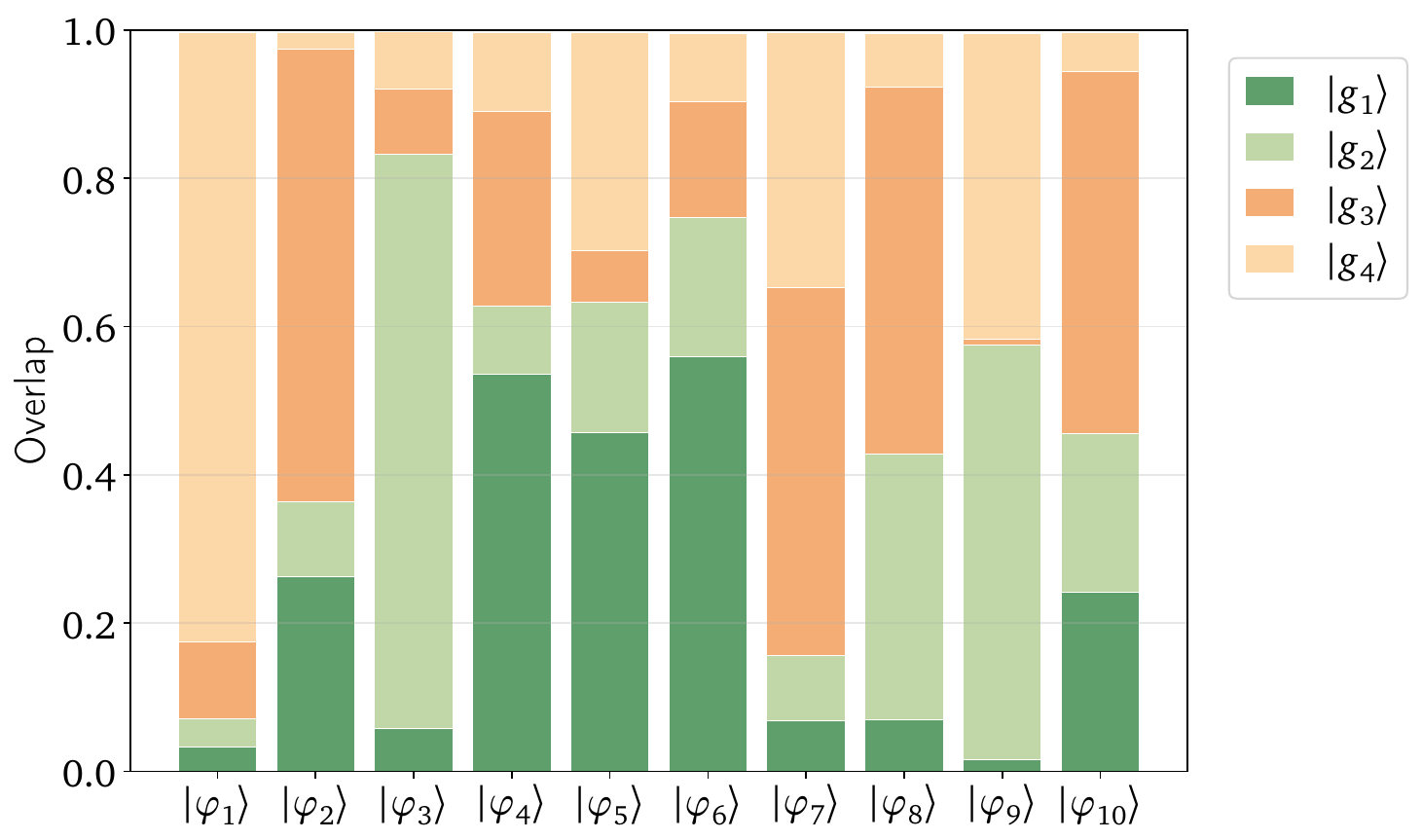}
  \caption{Overlaps of 10 representative generated states $\{|\psi_i\rangle\}_{i=1}^{10}$ with the exact AKLT ground-state basis $\{|g_i\rangle\}_{i=1}^{4}$ at system size $N=10$. Each column shows the decomposition of a generated state in the exact basis. The first 4 generated states form an approximately orthogonal subset.}
  \label{fig:app-aklt-generate_states}
\end{figure}

Fig.~\ref{fig:app-aklt-generate_states} shows the overlap decomposition of 10 representative generated states in the exact AKLT ground-state basis. The overlap patterns are visibly distinct from one sample to another, indicating that the diverse generated states correspond to different linear combinations of the exact basis states, rather than repeated copies of a few representative directions. In particular, the first four generated states, $|\psi_1\rangle$–$|\psi_4\rangle$, form an approximately orthogonal subset, with mutual overlaps below $0.08$. This provides direct sample-level evidence that the generated ensemble already contains a basis-resolving subset for the full AKLT ground space.

\begin{figure}[htbp]
  \centering
  \begin{subfigure}[t]{0.23\textwidth}
    \centering
    \includegraphics[width=\linewidth]{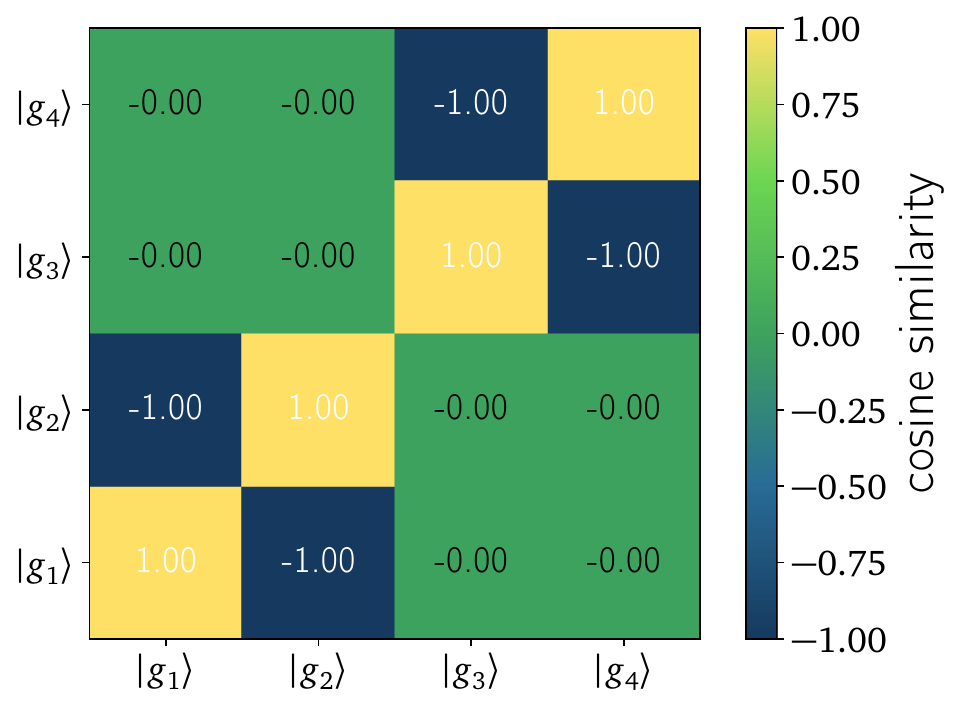}
    \caption{True states ($N=10, \bm{l}^{(1)}$)}
    \label{fig:aklt-true-l1}
  \end{subfigure}
  \hfill
  \begin{subfigure}[t]{0.23\textwidth}
    \centering
    \includegraphics[width=\linewidth]{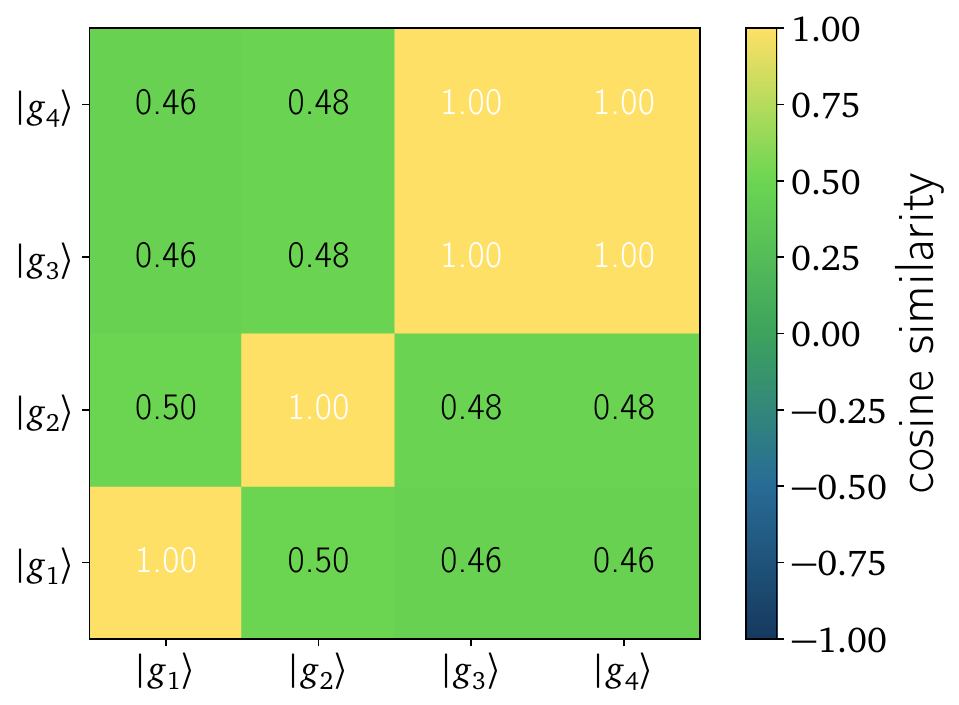}
    \caption{True states ($N=10, \bm{l}^{(2)}_{\text{edge}}$)}
    \label{fig:aklt-true-l2}
  \end{subfigure}

  \vspace{0.1mm}

  \begin{subfigure}[t]{0.23\textwidth}
    \centering
    \includegraphics[width=\linewidth]{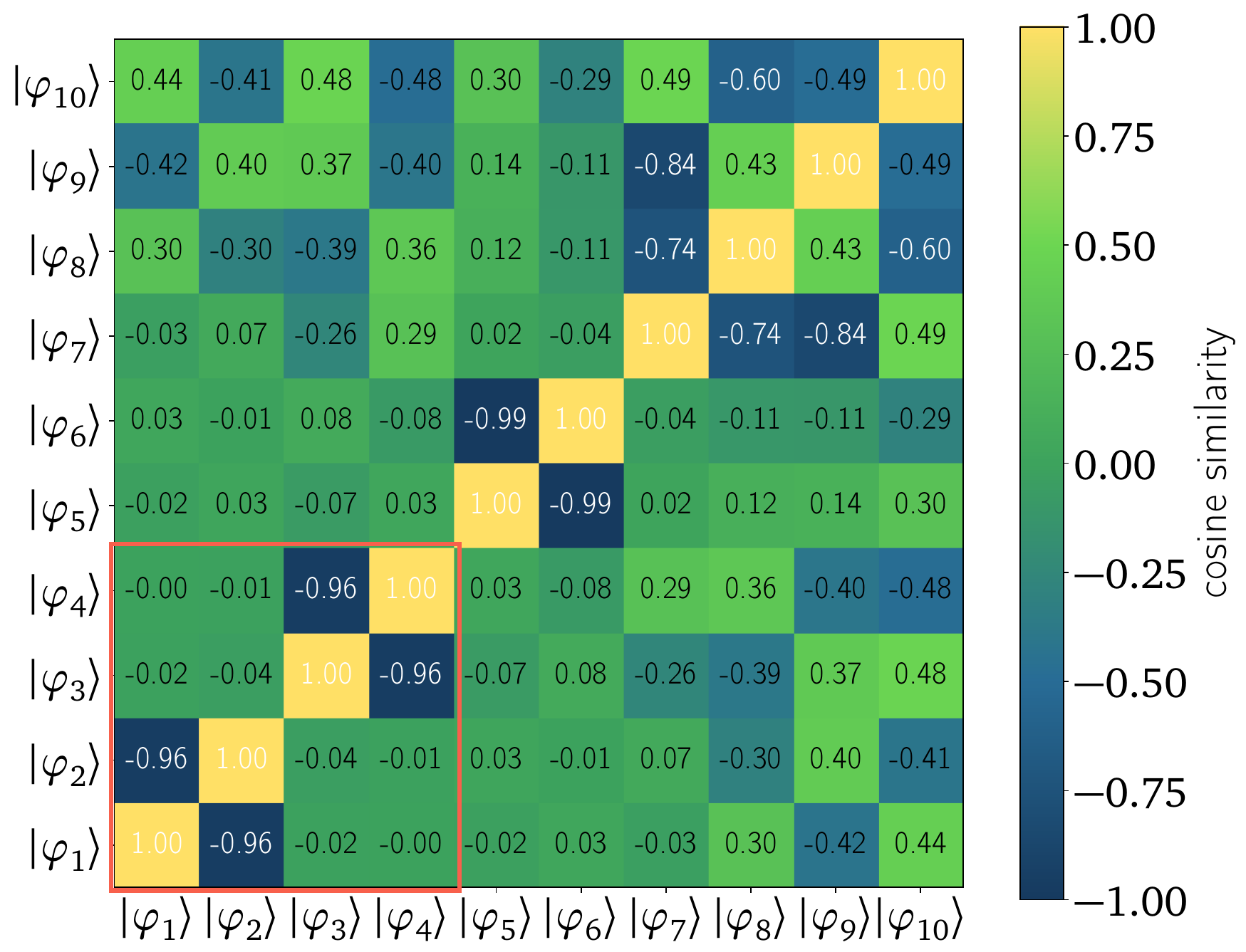}
    \caption{Generated states ($N=10, \bm{l}^{(1)}$)}
    \label{fig:aklt-gen-l1}
  \end{subfigure}
  \hfill
  \begin{subfigure}[t]{0.23\textwidth}
    \centering
    \includegraphics[width=\linewidth]{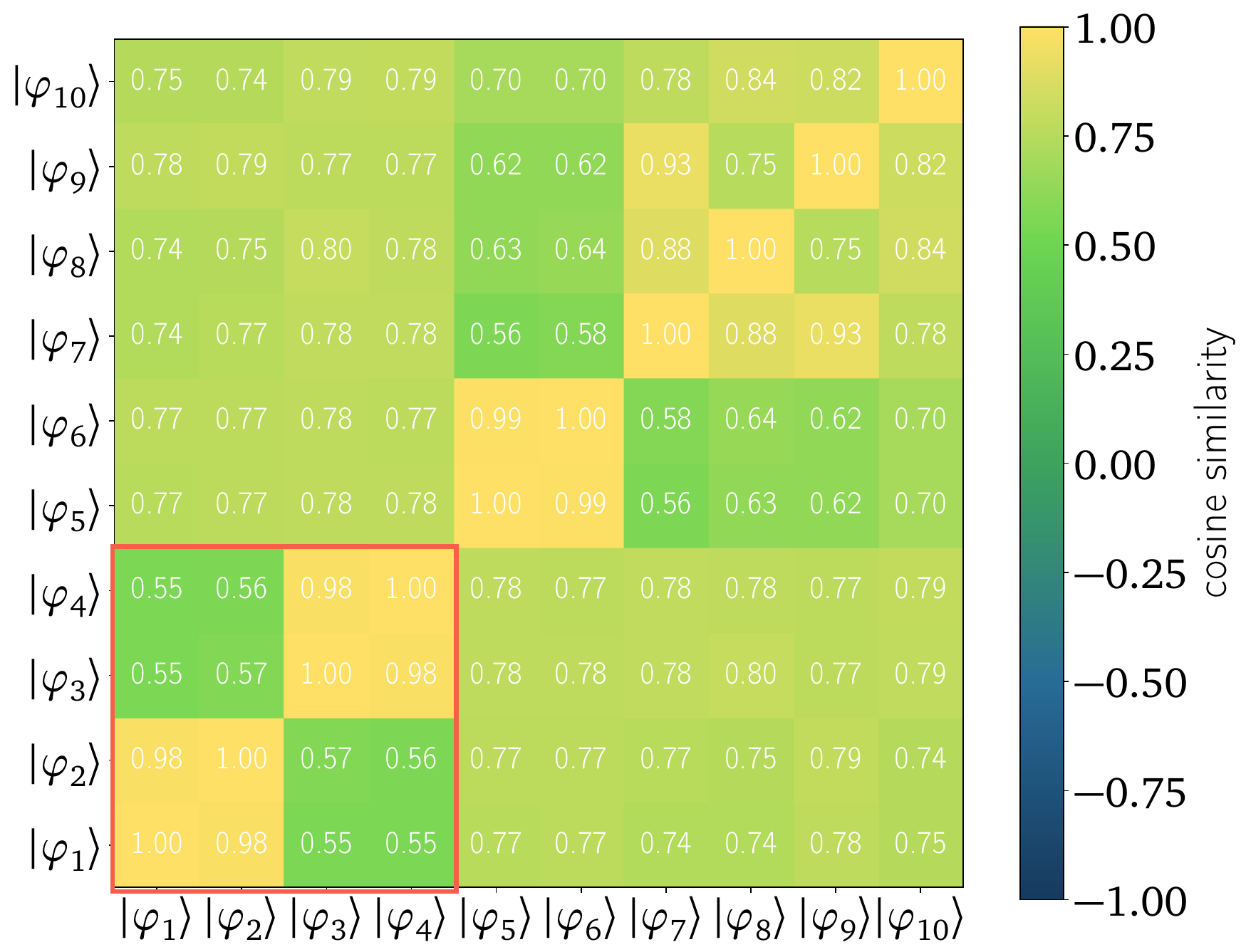}
    \caption{Generated states ($N=10, \bm{l}^{(2)}_{\text{edge}}$)}
    \label{fig:aklt-gen-l2}
  \end{subfigure}

  \caption{Cosine-similarity matrices for the AKLT model encoded in an $N=10$ qubit circuit. The upper row shows the exact ground-state basis $\{|g_i\rangle\}_{i=1}^{4}$, and the lower row shows 10 representative generated states $\{|\psi_i\rangle\}_{i=1}^{10}$. Similarities are computed from the one-body features $\bm{l}^{(1)}$ and the edge-restricted two-body correlators $\bm{l}^{(2)}_{\text{edge}}$. The red boxes highlight the subset of generated states forming an approximately orthogonal basis whose feature-space structure matches that of the exact states.}
  \label{fig:aklt-cosine-comparison}
\end{figure}

To probe the structure of the generated ensemble in local-observable feature space, we next compare the cosine-similarity matrices of the exact and generated states. For this purpose, we use both the one-body feature vector $\bm{l}^{(1)}$ and the edge-localized two-body feature vector $\bm{l}^{(2)}_{\text{edge}}$, defined in Eqs.~(\ref{eq:one/twocorre}) and (\ref{eq:aklt-two-edge}) of the main text. The exact ground states show that these two feature sets carry complementary information. In the one-body feature space $\bm{l}^{(1)}$, the states are either nearly orthogonal ($\bm{l}^{(1)} \approx 0$) or strongly anticorrelated ($\bm{l}^{(1)} \approx -1$), indicating that the local magnetization profile already distinguishes different edge-spin configurations (Fig.~\ref{fig:aklt-true-l1}). However, the edge-restricted two-body correlations $\bm{l}^{(2)}_{\text{edge}}$ capture a different aspect of the structure. Most off-diagonal entries lie in a narrow positive range, while the pair $|g_3\rangle$ and $|g_4\rangle$ becomes nearly indistinguishable in this representation ($\bm{l}^{(2)}_{\text{edge}} \approx 1$, Fig.~\ref{fig:aklt-true-l2}). Together, these two feature sets provide a compact and characteristic description of the AKLT ground space.

The same analysis applied to the generated states yields closely matching patterns in Figs.~\ref{fig:aklt-gen-l1} and \ref{fig:aklt-gen-l2}. In particular, within the identified approximately orthogonal subset, the generated states reproduce the same separation mechanism seen in the exact basis. Certain pairs are sharply distinguished in the one-body feature space through strong anticorrelation, while the edge-localized two-body correlators retain the corresponding boundary correlation structure. The close agreement between the upper and lower rows of Fig.~\ref{fig:aklt-cosine-comparison} shows that the generative quantum circuit captures not only the correct AKLT ground-space span, but also the edge-state structure that underlies its degeneracy.

\subsection{Spin-1 XXZ model}
\label{app:xxz}

We consider the spin-1 XXZ chain at the first-order transition point $\Delta=-1$, which is defined as $H_{\mathrm{XXZ}} = \sum_{i=1}^{N-1} \left( S_i^x S_{i+1}^x + S_i^y S_{i+1}^y + \Delta  S_i^z S_{i+1}^z \right)$ ($S_i^{\alpha}$ are spin-1 operators at site $i$ and $\alpha = x,y,z$), where the ground space forms a ferromagnetic multiplet. In this appendix, we examine how representative generated states are distributed with respect to the exact ground-state basis, and whether the generated ensemble reproduces the feature-space structure of the exact ground space.

\begin{figure}[htbp]
    \centering
    \includegraphics[width=0.7\linewidth]{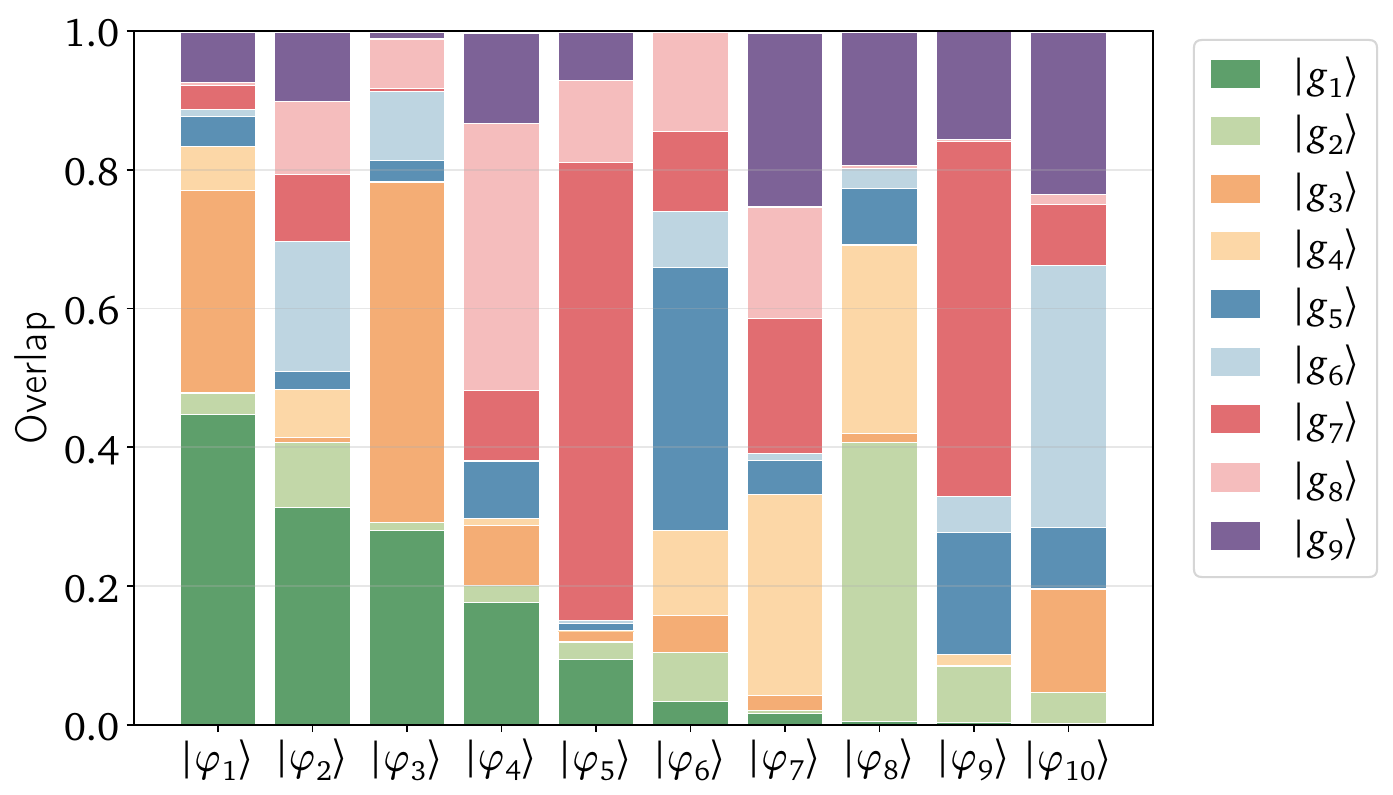}
  \caption{Overlaps of 10 representative generated states $\{|\psi_i\rangle\}_{i=1}^{10}$ with the exact spin-1 XXZ ground-state basis $\{|g_i\rangle\}_{i=1}^{9}$ at system size $N=8$. Each column shows the decomposition of a generated state in the exact basis. The distinct overlap patterns indicate that the generated states correspond to different linear combinations of the exact ground-state basis vectors.}
  \label{fig:app-xxz-generate_states}
\end{figure}

Fig.~\ref{fig:app-xxz-generate_states} shows the overlap decomposition of 10 representative generated states in the exact XXZ ground-state basis. The overlap patterns vary substantially from one sample to another, showing that the representative generated states probe different magnetization-resolved directions within the ferromagnetic multiplet. This provides direct sample-level evidence that the generated ensemble explores the full ferromagnetic multiplet in a balanced way.
\begin{figure}[htbp]
  \centering
  \begin{subfigure}[t]{0.23\textwidth}
    \centering
    \includegraphics[width=\linewidth]{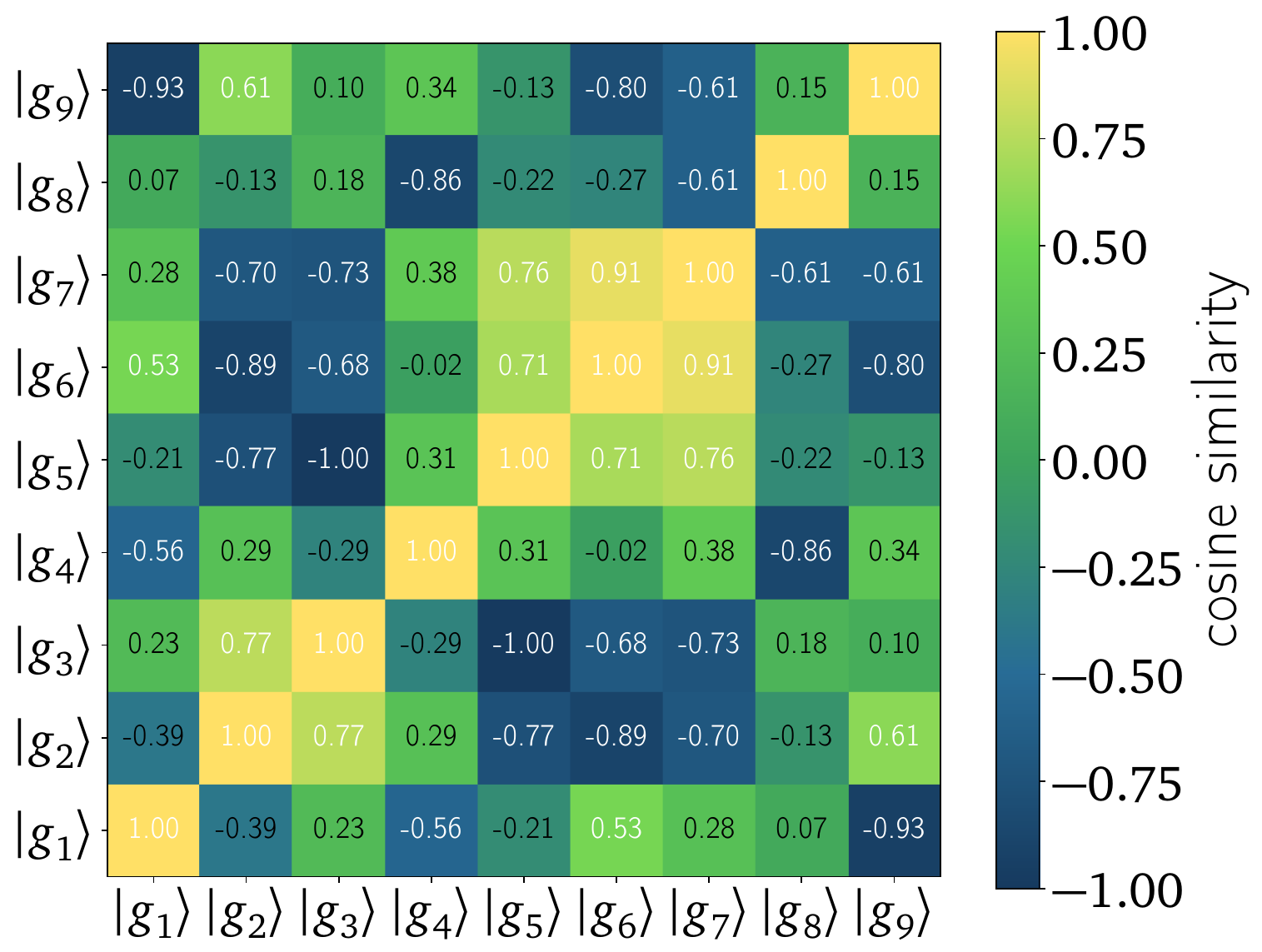}
    \caption{True states ($N=8, \bm{l}^{(1)}$)}
    \label{fig:xxz-true-l1}
  \end{subfigure}
  \hfill
  \begin{subfigure}[t]{0.23\textwidth}
    \centering
    \includegraphics[width=\linewidth]{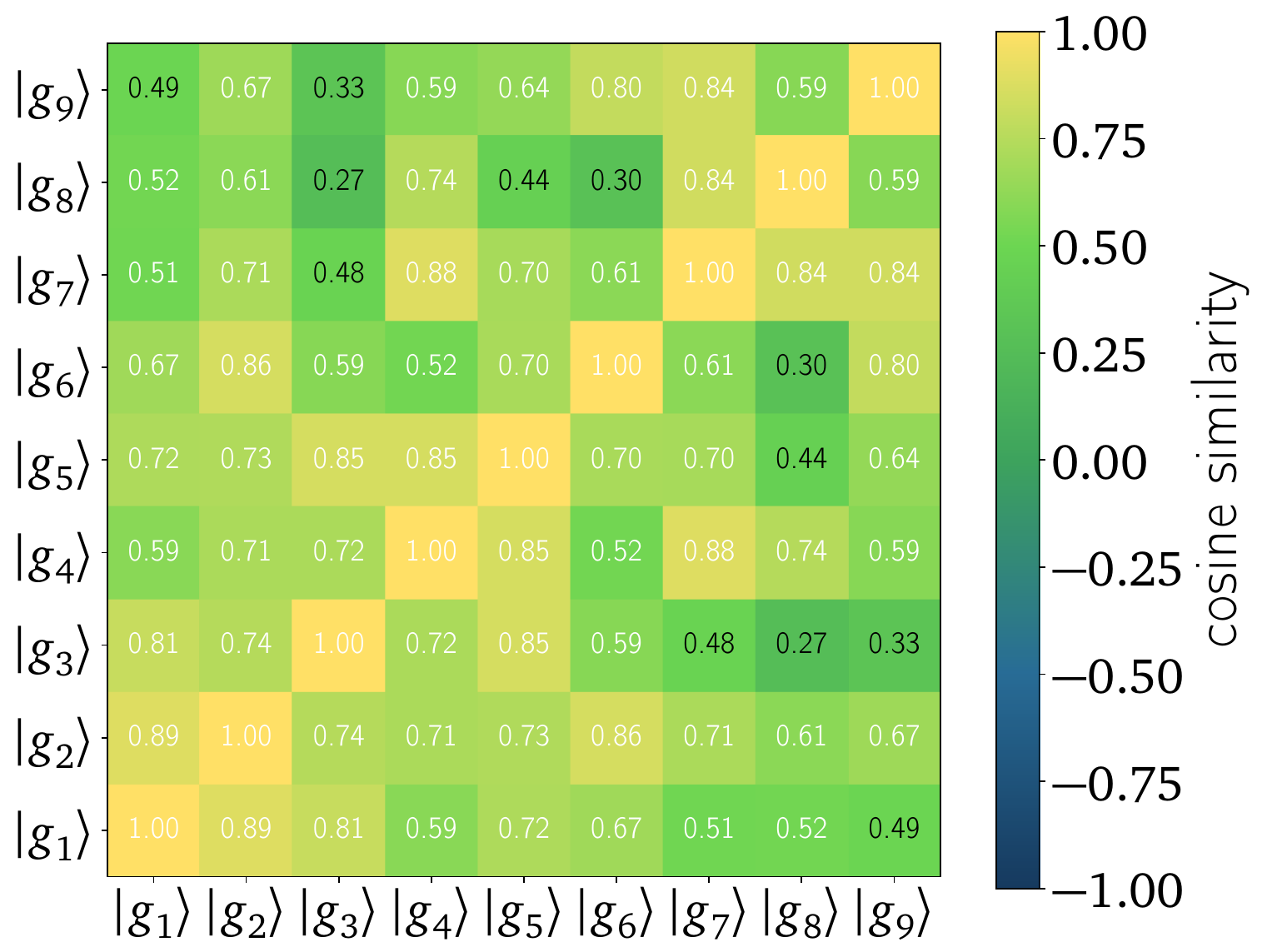}
    \caption{True states ($N=8, \bm{l}^{(2)}$)}
    \label{fig:xxz-true-l2}
  \end{subfigure}

  \vspace{0.3cm}

  \begin{subfigure}[t]{0.23\textwidth}
    \centering
    \includegraphics[width=\linewidth]{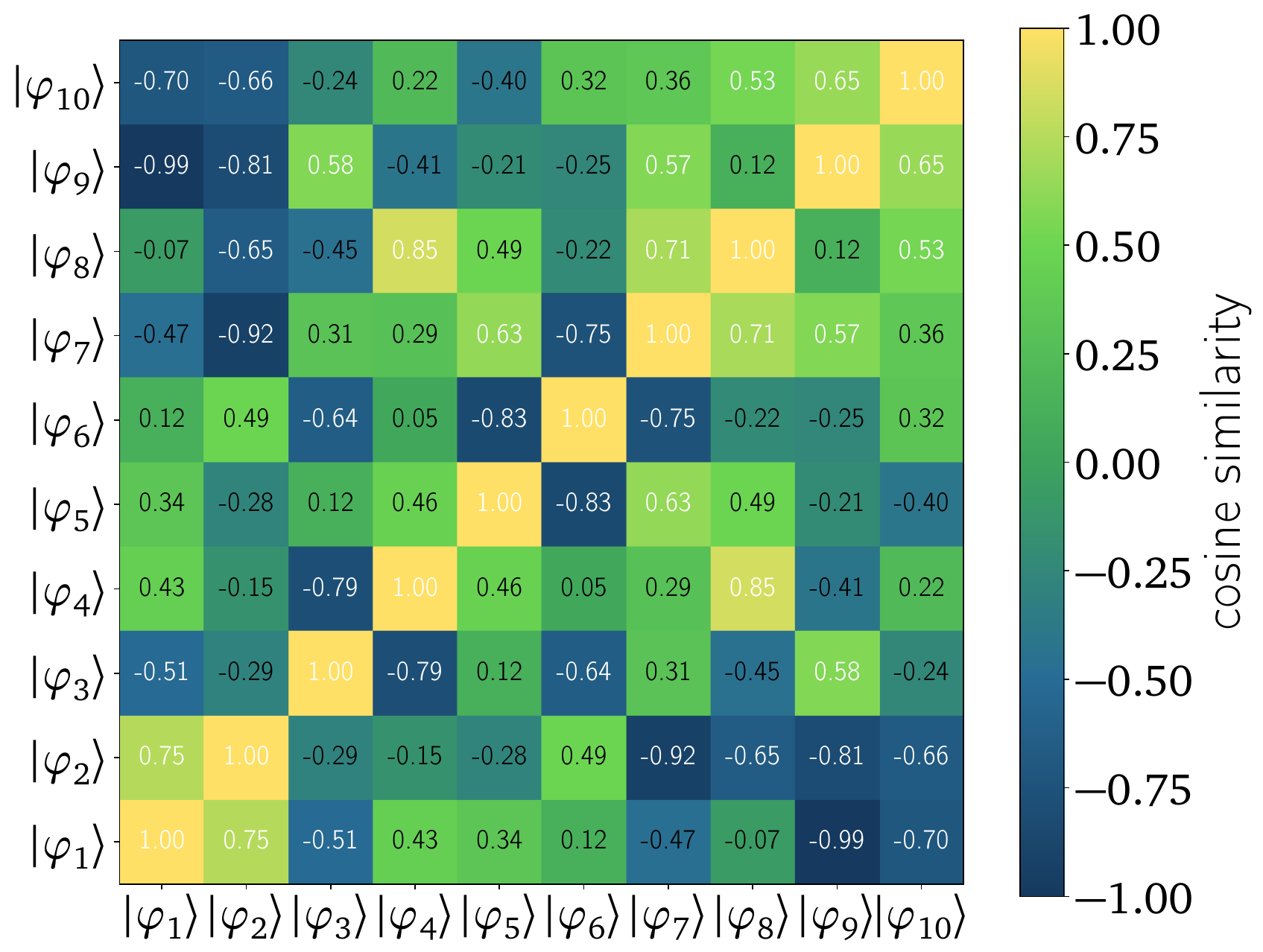}
    \caption{Generated states ($N=8, \bm{l}^{(1)}$)}
    \label{fig:xxz-gen-l1}
  \end{subfigure}
  \hfill
  \begin{subfigure}[t]{0.23\textwidth}
    \centering
    \includegraphics[width=\linewidth]{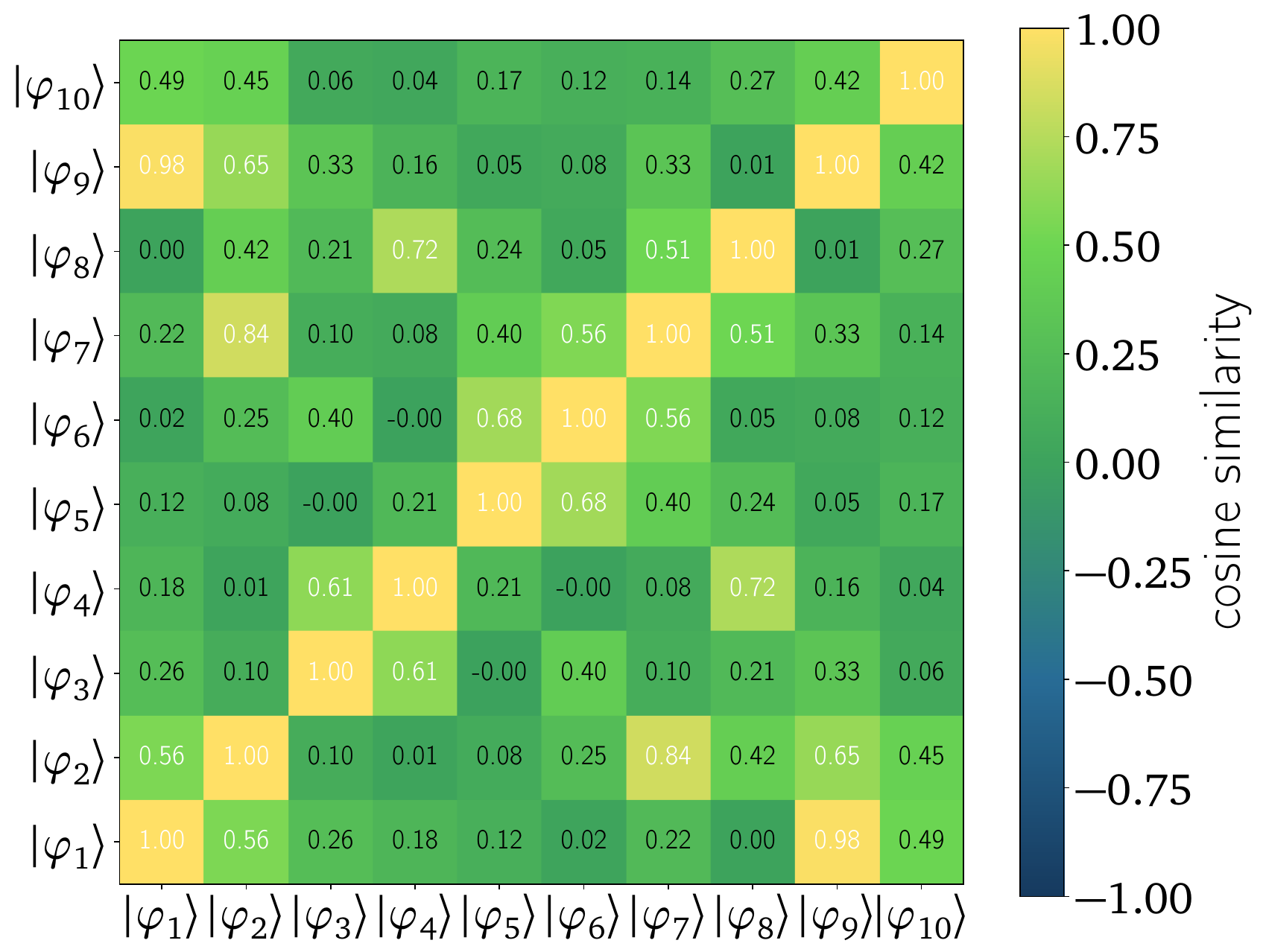}
    \caption{Generated states ($N=8, \bm{l}^{(2)}$)}
    \label{fig:xxz-gen-l2}
  \end{subfigure}

  \caption{Cosine-similarity matrices for the spin-1 XXZ model encoded in an $N=8$ qubit circuit. The upper row shows the exact ground-state basis $\{|g_i\rangle\}_{i=1}^{9}$, and the lower row shows 10 representative generated states $\{|\psi_i\rangle\}_{i=1}^{10}$. Similarities are computed from the one-body expectation values $\bm{l}^{(1)}$ and the nearest-neighbor two-body correlators $\bm{l}^{(2)}$. The generated states reproduce the same broad one-body variation and positive two-body correlation structure as the exact ferromagnetic ground multiplet.}
  \label{fig:xxz-cosine-comparison}
\end{figure}

To understand the structure of the exact ground space in local-observable feature space, we next examine the cosine-similarity matrices of the exact XXZ ground states. The one-body feature matrix $\bm{l}^{(1)}$, shown in Fig.~\ref{fig:xxz-true-l1}, exhibits a broad range of values determined by the distinct local magnetization patterns of the ferromagnetic multiplet. In particular, the similarity values range from strong anticorrelation, such as $\bm{l}^{(1)}\approx -1$ between $|g_3\rangle$ and $|g_5\rangle$, to strong positive correlation, such as $\bm{l}^{(1)}\approx 0.91$ between $|g_6\rangle$ and $|g_7\rangle$. This shows that the one-body observables sharply distinguish different magnetization sectors. By contrast, the two-body correlators in Fig.~\ref{fig:xxz-true-l2} remain strictly positive, with off-diagonal similarities concentrated in the range $[0.27,0.89]$. These two-body features therefore reflect the shared ferromagnetic coherence of the multiplet, while the one-body features resolve the differences between its individual states. Taken together, $\bm{l}^{(1)}$ and $\bm{l}^{(2)}$ provide a compact but complete feature-space description of the degenerate XXZ ground space.

We now compare these exact patterns with those of the generated ensemble. As shown in Figs.~\ref{fig:xxz-gen-l1} and \ref{fig:xxz-gen-l2}, the cosine-similarity matrices of the generated states closely reproduce the characteristic structure of the exact ground space. In the one-body feature space, the generated states retain the broad variation associated with different local magnetization patterns. In the two-body feature space, they preserve the uniformly positive correlation structure characteristic of the ferromagnetic multiplet. The close agreement between the exact and generated cosine-similarity matrices therefore shows that the local-observable-guided training captures not only the correct ground-space span, but also the internal feature-space geometry of the XXZ ground states.

\end{document}